\documentclass[a4paper,12pt]{article}
 \usepackage[latin1]{inputenc}
\usepackage[T1]{fontenc}
\usepackage{amsmath}
\usepackage{amsfonts}
\usepackage{amssymb}
\usepackage{color}
\usepackage{graphicx}
\usepackage{caption}
\usepackage{comment}
\usepackage{subcaption}%

  \def\hD{{\hat D}}
 \newcommand{\be}{\begin{equation}}
	 \newcommand{\ee}{\end{equation}}
	 \newcommand{\ba}{\begin{eqnarray}}
		 \newcommand{\ea}{\end{eqnarray}}
		 
		   \newcommand{\bea}{\begin{eqnarray}}
			 \newcommand{\eea}{\end{eqnarray}}
 \newcommand{\nn}{\nonumber}

\newcommand{\nl}{\nonumber \\} 
\newcommand{\beq}{\begin{equation}}
\newcommand{\eeq}{\end{equation}}
\newcommand{\p}{\partial}
\def\a{\alpha}
\def\b{\beta}
\def\le{\left(}
\def\ri{\right)}
\def\tx{\tilde{x}}
\def\Tr{{\rm Tr}}
\def\im{{\rm Im}}
\def\re{{\rm Re}}
\def\ta{\tilde{\alpha}}
\def\tx{{\tilde{x}}}
\def\hD{{{\hat{D}}}}
\def\hH{{{{\hat{H}}^\omega}}}
\def\hpi{{{{\hat{\pi}}}^\omega}}
\def\hpiz{{{{\hat{\pi}}}_0}}
\def\hHl{{{{\hat{H}}^{\omega,1}}}}
\def\hHq{{{{\hat{H}}^{\omega,2}}}}
\def\hpiq{{{{\hat{\pi}}}^{\omega,2}}}
\def\o{\omega}

\begin{document} 
\title{The holographic non-abelian vortex}
\author{Gianni Tallarita$^{a}$, Roberto Auzzi $^{b,c}$ and  Adam Peterson$^{d}$  \\\\
  $^{a}$ {\it \small Departamento de Ciencias, Facultad de Artes Liberales,
 Universidad Adolfo Ib\'a\~nez,}\\ 
 { \it \small  Santiago 7941169, Chile,}\\\\
   $^{b}$ {\it \small  Dipartimento di Matematica e Fisica,  Universit\`a Cattolica
del Sacro Cuore, }\\
{\it \small  Via Musei 41, 25121 Brescia, Italy, } 
\\\\
$^c$ {\it \small INFN Sezione di Perugia,  Via A. Pascoli, 06123 Perugia, Italy, }\\\\
  $^{d}$ {\it \small Department of Physics, University of Toronto, }\\ 
{\it \small Toronto ON, Canada.} 
}
\date{\hfill}

\maketitle
\begin{abstract}

We study a fully back-reacted non-abelian vortex solution
in an extension of the holographic superconductor setup.
The thermodynamic properties of the vortex are computed.
We show that, in some regime of parameters, the non-abelian vortex solution
has a lower free energy than a competing abelian vortex solution. 
The solution is dual to a finite-temperature
perturbed conformal field theory with a topological defect, 
on which operators related to the Goldstone modes of a spontaneously broken symmetry
are localized. We compute numerically  the retarded Green function of these operators
and we find, in the classical approximation in the bulk, a gapless $\mathbb{CP}^1$ excitation
on the vortex world line.

\end{abstract}

\section{Introduction}
 
 \quad The  non-abelian vortex \cite{Hanany:2003hp,Auzzi:2003fs} 
 is characterized by the presence of orientational
 internal zero modes localized on the vortex.
 These modes are generated by a symmetry, which is unbroken 
 in the bulk and spontaneously broken in the core of the vortex.
 The dynamics of these orientational modes is described by an effective
 $\mathbb{CP}^{N}$ sigma model localized on the vortex worldsheet.
 
One of the most important motivations to study these objects is that they may give
us precious insights on the challenging problem of quark confinement.
 The dual mechanism of color confinement
  \cite{'tHooft:1981ht,Mandelstam:1974pi} suggests the existence 
  of a confining superconductor-like vortex between quark charges. 
  Indeed, from the major breakthrough by Seiberg and Witten \cite{Seiberg:1994rs}, we know that
the Abrikosov-Nielsen-Olesen (ANO) vortex in an effective dual  abelian Higgs model
  realizes confinement in softly broken $\mathcal{N}=2$ Super Yang-Mills theory.
  On the other hand, the abelian model fails to reproduce several features
  of realistic confinement: in particular it gives an extra multiplicity in the hadron spectrum  
  \cite{Vainshtein:2000hu,Yung:2001cz},  which is not observed in nature. 
  Indeed, the ANO type cofinement arises from 
  the breaking of a $U(1)$ gauge symmetry, which would imply
     some degree of abelianization of $SU(3)$ QCD theory which is not necessarily realistic.
   It is then interesting to study  alternative kinds of confining vortex strings, such as the 
   non-abelian vortex, which realizes confinement in some vacua of 
 $\mathcal{N}=2$ SQCD \cite{Carlino:2000uk}.
In several $\mathcal{N}=2$ gauge theories the non-abelian vortex leads 
to a powerful correspondence between  theories in different dimensions:
 the BPS spectrum of the two-dimensional effective sigma model coincides with the spectrum
  of monopoles in the four-dimensional gauge theory  \cite{Dorey:1999zk,Hanany:2004ea,Shifman:2004dr}.
  In the supersymmetric case, this gives a nice  setting
  where one can study the physics of strongly-coupled gauge theories 
 in a controlled way.
    
Non-abelian vortices have been heavily investigated in several contexts, 
e.g. \cite{Shifman:2012vv,Shifman:2014oqa,Peterson:2014nma,Peterson:2015tpa,Eto:2006dx,Forgacs:2016dby},
for reviews see \cite{Tong:2005un,Eto:2006pg,Konishi:2008vj,Shifman:2012zz,shifman2}.  
In general, the worldsheet dynamics of a non-abelian vortex
   is itself strongly coupled at low energy
   (due to the asymptotic freedom of the  $\mathbb{CP}^{N}$ sigma model)
   and   we quickly lose analytic control outside the comfort zone 
   of supersymmetric theories.

Holography gives us an interesting theoretical laboratory to explore strongly coupled systems in a calculable setting.
Even if no explicit examples of real world materials described by a gravity dual have been found so far in a lab,
gravity duals give us examples of theoretically consistent physical systems where no quasi-particle description is available.
This is one of the situations where the traditional field-theoretical tools fail.
Holographic superconductors \cite{Gubser:2008px,Hartnoll:2008vx,Hartnoll:2008kx} 
 provide an interesting class of the exotic phases of matter that have been recently studied in holography. These 
 systems are very different from the conventional superconductors
 described with the effective field theory framework, which are generically described
 by a small number of degrees of freedom, i.e. the spontaneous breaking of a $U(1)$ gauge symmetry.
Holographic superconductors instead possess a large number of gapless degrees of freedom,
obtained by deforming a conformal field theory (CFT) 
 by relevant operators. The non-trivial interaction between Goldstone bosons and CFT
 gives a rich dynamics which has been intensively studied in the last few years. 
 This program includes  abelian vortices in holographic superconductors and superfluids as an interesting example,
 see  \cite{Albash:2009iq,Montull:2009fe,Maeda:2009vf,Keranen:2009re,Tallarita:2010vu,Iqbal:2011bf,Dias:2013bwa}.

In the quest of investigating the several possible realisations of strongly coupled systems,
it is then natural to explore holographic models of non-abelian vortices.
Non-abelian vortices were previously studied in gravity duals
of strongly-coupled gauged theories in mass-deformed $\mathcal{N}=4$ 
and ABJM \cite{Aharony:2008ug} theories,
 always in the probe  approximation,
  i.e. as probe D-brane in the background geometry \cite{Auzzi:2008ep,Auzzi:2009yw,Auzzi:2009es}.
 Considering D-brane back-reaction could give a top-down realisation of non-abelian vortices
 in AdS/CFT, but this is a hard problem. 
 
 Here we take a pragmatical approach and we consider a
 bottom/up prospective, in order to build a non-abelian vortex
  in AdS with a minimal amount of ingredients. 
 Inspired by the flat space vortex string construction 
 studied in \cite{Witten:1984eb,Shifman:2012vv},
a holographic model of non-abelian vortex was proposed 
and studied in \cite{Tallarita:2015mca}.
The purpose of this paper is to  include the effect of the gravity back-reaction  and also
to further investigate the issue of non-abelian zero modes, which corresponds
to Goldstone modes localized on the vortex world volume. 
Therefore, this is the first realisation of a fully back-reacted non-abelian vortex in a holographic setup. \newline

The paper is organized as follows: in section \ref{sect2}
we review the abelian vortex in the holographic superconductor model studied in 
\cite{Dias:2013bwa}. In section \ref{sect3} we introduce the non-abelian
vortex set-up (which is a variant of the one studied in \cite{Tallarita:2015mca})
and we find the vortex solution, including the gravity back-reaction.
A free energy calculation (which is necessary
in order to establish if  the non-abelian vortex solution is dynamically preferred)
is performed in section \ref{sect4}. The orientational zero modes and the retarded
two-points function of the Goldstone bosons localized
on the vortex worldsheet are studied in  section \ref{sect5}.
We conclude in section \ref{sect-conclu}.

\section{Gravitating abelian vortex}
\label{sect2}

In the context of holographic superconductivity \cite{Hartnoll:2008kx}, 
the topic of abelian vortex solutions have been studied in several papers, e.g.
  \cite{Albash:2009iq,Montull:2009fe,Maeda:2009vf,Tallarita:2010vu}.  
In these studies, the back-reaction of the matter fields 
on the gravitational sector is neglected, and so only a
 limited parameter space of the full solution is considered.
 In particular they are restricted to temperatures far from $T=0$, 
 where the back-reaction can no longer be ignored.

A fully back-reacted gravitational system which is dual to a three dimensional theory 
containing a vortex of the Abrikosov type was first constructed in \cite{Dias:2013bwa}.  
We dedicate this section to a lightning review of this system and
 its notation as it is very important for this paper.  The system introduced 
 in \cite{Dias:2013bwa} is a gravitational Maxwell-Higgs system of the form:
\be\label{sano}
S_{ANO} = \frac{1}{16\pi G_N}\int d^4x \sqrt{-g}\left[R+\frac{6}{L^2}
-\frac{1}{2}F_{\mu \nu}F^{\mu \nu }-2(D_\mu \phi)(D^\mu \phi)^{\dagger}-2V(|\phi|^2)\right],
\ee
where
\be
F_{\mu \nu} = \partial_\mu A_\nu -\partial_\nu A_\mu,
\, , \qquad
D_\mu \phi = \partial_\mu \phi - iq A_\mu \phi.
\ee
In our notation $D_\mu$ denotes in general a combination of the gravity
and $U(1)$ gauge covariant derivatives.
This system has a $U(1)$ gauge symmetry, spontaneously broken by a quartic potential of the form
\be
\label{pot}
V(|\phi|^2) = -\frac{2}{L^2}|\phi|^2\left(1-\frac{1}{2  }|\phi|^2\right).
\ee
The potential has two local extremal points, 
with different $V$ and mass for the scalar $\phi$:
\bea
|\phi| &=& 0  \, , \qquad V=0 \, , \qquad m^2_\phi=-\frac{2}{L^2} \, , \nl
|\phi| &=& 1 \, , \qquad V=-\frac{1}{L^2}  \, , \qquad m^2_\phi=\frac{4}{L^2} \, .
\eea
We will focus on the first of these  AdS vacua:
\be
ds^2 = \frac{L^2}{z^2}(-dt^2+dz^2+dr^2+r^2d\theta^2) \, ,
\label{empty-ads}
\ee
for which at large $z$ the field $\phi$ has the following expansion
\be
\phi = \alpha z^{\Delta_1}+ \beta z^{\Delta_2} +... \, , \qquad \Delta_1=1 \, , \qquad \Delta_2=2 \, .
\label{bou-phi}
\ee
For generic mass $m^2_\phi$ ,  the  dimensions $\Delta_i$ are the solutions of $m^2_\phi L^2=\Delta (\Delta-3)$,
which comes from the Klein-Gordon equation in curved AdS space.

The general setup involving a holographic phase transition
 is to find a phase of the gravitational system where a black-hole forms scalar hair.  
 This usually involves the presence of a chemical potential in the dual theory. 
 The reason being that without it the dual system is scale invariant and every non-zero temperature is equivalent.  
 The chemical potential is introduced by using the temporal component 
 of the gauge field $A_0$ in the bulk, whose asymptotic 
 boundary condition is then dual to the chemical potential. 
 The authors of \cite{Dias:2013bwa}  (using results from \cite{Faulkner:2010gj}) 
however achieve the same mechanism without using a chemical potential, 
instead they impose Robin boundary conditions on the scalar field at the boundary. 
  In the usual case, following the holographic dictionary, one imposes that either $\alpha$ or $\beta$ vanish, 
  as if one assumes any of them to describe the dual condensate the other can be understood as its source 
  and one has to abandon the idea of a spontaneous phase transition.  
  Therefore, for $\alpha=0$ or $\beta =0$ the scalar field $\phi$ is dual to an
   operator $\mathcal{O}$ with conformal dimension $2$ or $1$.

Imposing a Robin condition of the form
\be\label{bc}
\beta = \kappa \alpha
\ee
is dual to introducing a relevant double-trace  operator in the dual theory potential 
\cite{Witten:2001ua,Berkooz:2002ug} of the form 
 \beq
 \Delta \mathcal{V}=  \kappa \,  \mathcal{O}^\dagger\mathcal{O} \, ,
 \eeq
  with a positive coefficient.
  For negative values of $\kappa$ this term induces a condensation
   of the $\mathcal{O}$ operator in the dual theory, 
   determined  by the new scale $\kappa$.

With this boundary condition describing a holographic phase transition the authors of \cite{Dias:2013bwa} 
solved the full system of the equation of motion:
\be\label{eqs}
G_{\mu \nu}=R_{\mu \nu}+\frac{3}{L^2} g_{\mu \nu}-\left[(D_\mu \phi)(D_\nu \phi)^{\dagger}
+(D_\nu \phi)(D_\mu \phi)^{\dagger}+g_{\mu \nu}V(|\phi|^2)+F_\mu^\sigma F_{\sigma \nu }
-\frac{g_{\mu \nu}}{4}F^{\rho \sigma}F_{\rho \sigma }\right]=0,
\ee
\be\nn
D_\mu F^{\mu \nu}=i q[(D^\nu \phi)\phi^{\dagger}-(D^\nu \phi)^{\dagger}\phi],
\ee
\be\nn
g^{\mu \nu}D_\mu D_\nu \phi-V'(|\phi|^2)\phi=0,
\ee
which come from the action in eq.~(\ref{sano}).
The following ansatz, which is the most general one consistent with cylindrical symmetry, is used:
\bea
\label{metric}
ds^2 &=& \frac{L^2}{y^2}\left\{ -Q_1 y_+^2(1-y^3)dt^2+\frac{Q_2}{1-y^3}dy^2
 \right. \nn\\
&& \left.  +\frac{y_+^2Q_4}{(1-x)^4}(dx+xy^2(1-x)^3Q_3 dy)^2
+  \frac{y_+^2 Q_5 x^2}{(1-x)^2}d\theta^2 \right\} \, ,
\eea
\be
\phi = y e^{i n \theta}x^n Q_6, \quad A_\theta = Lx^2 Q_7.
\ee
In eq.~(\ref{metric}) the radial AdS coordinate is
$y \in [0,1]$, with $y=0$ being the conformal boundary and $y=1$ 
the horizon.  
The form of the metric in eq.~(\ref{metric}) involves the following change
of variables from the usual cylindrical coordinate $r$  (defined from $0 \leq r \leq \infty$) 
transverse to the vortex axis of symmetry:
\be
r = \frac{x}{1-x}, \qquad  x \in [0,1] \, .
\ee
Empty AdS corresponds to: 
\beq
Q_1=\frac{1}{1-y^3} \, , \qquad Q_2=1-y^3 \, , \qquad Q_3=0 \, , \qquad
Q_4=1 \, , \qquad Q_5=1 \, ,
\eeq
and the black brane solution (without  any scalar field) instead corresponds to :
\beq
Q_1 = Q_2 = Q_4 = Q_5 = 1, \quad  Q_3=0.
\eeq

Using this ansatz in the equations of motion eq.(\ref{eqs}) leads to a complicated 
 set of  coupled non-linear pdes for the $Q_i(x,y)$
 fields which we choose not to display for simplicity, 
but which we plan to solve anyway.
  
The temperature is:
\be\label{T2}
T= \frac{3 y_+}{4\pi}.
\ee
This definition relies on the fact that $Q_1(x,1) = Q_2(x,1)$ which is part 
of a boundary condition enforcement as we discuss in the next section.

As they stand however the equations are not elliptic and 
are difficult to solve numerically.
 The authors of \cite{Dias:2013bwa}  adopted the DeTurck method in order to make the equations elliptic,
  this method is explained in detail in \cite{Dias:2015nua}. 
  In practice, one must solve a modified version of Einstein's equations, called Einstein-DeTurck equations, of the form
\be
G_{\mu \nu}- D_{(\mu }\xi_{\nu)}=0,
\ee
where $G_{\mu \nu}$ is given in equation (\ref{eqs}) and 
\be
\xi^\mu = g^{\rho \sigma }\left(\Gamma_{\rho \sigma }^\mu(g)-\bar{\Gamma}_{\rho \sigma }^\mu (\bar{g})\right).
\ee
Here $\bar{g}$ is a reference metric chosen to have the same asymptotic conditions as the original metric eq.(\ref{metric}). This new system of equations is elliptic and can be solved by standard numerical methods. However, one must guarantee that the solution found has a vanishing DeTurck vector $\xi^a\xi_a =0$, otherwise the solution is not a solution of the original Einstein Matter system, but rather a Ricci soliton. For this case the reference metric is chosen to be the same line element as in eq.(\ref{metric}) with
\be
Q_1 = Q_4=Q_5=1\, , \quad Q_3=0 \, ,
\, \qquad
Q_2=1-\tilde{\alpha} y(1-y) \, ,
\label{alphatilde}
\ee
where $\tilde{\alpha}$ is a constant that we will be discussed in section \ref{sect4}.

\subsection{Boundary Conditions}

The boundary conditions on this 2d system are very important and they were discussed at length in \cite{Dias:2013bwa}:
\begin{itemize}

\item {$y=0$} 

Here we require that metric tends to the black brane solution, therefore
\be
Q_1 = Q_2 = Q_4 = Q_5 = 1, \quad  Q_3=0.
\ee
The boundary condition on the scalar field was already discussed, 
this is the previously mentioned condition eq.(\ref{bc}). 
With the $Q_i$ parameterization this amounts to a Robin boundary condition on $Q_6$
\be
\partial_y Q_6(x,0) = \frac{\kappa_1}{y_+} Q_6(x,0),
\ee
where $\kappa_1$ is related to $\kappa$ and to the  $\tilde{\alpha}$
parameter in the DeTurck reference metric eq.(\ref{alphatilde}), 
as we will make explicit later in eq. (\ref{kap}).
The boundary condition on the gauge field correspond to $Q_7(x,0)=0$ for a superfluid, 
while $\partial_y Q_7(x,0)=0$ for a superconductor.

\item  {$x=1$}

The conditions infinitely far away from the vortex core are not so simple.
 At this boundary we require the solutions to approach an equivalent 
 superconducting system without the vortex. 
 That is, we must solve the same system removing the any dependence on the spatial coordinate $x$.
  The corresponding solution will be a superconducting state, and this solution
   must be fed as an asymptotic boundary condition infinitely far away from the vortex core. 
   The corresponding boundary conditions for a superconducting phase are therefore
\bea
Q_1(1,y) &=& \tilde{Q}_1(y), \quad Q_2(1,y) = \tilde{Q}_2(y) , \quad  Q_3(x,1) = 0,
\nl
Q_4(1,y) &=& Q_5(1,y) = \tilde{Q}_3(y), \quad  Q_6(1,y) =\tilde{Q}_6(y)  , \quad  Q_7(1,y) = n/qL,
\eea
where $\tilde{Q}_i(y)$ are the spatially independent functions which solve the homogeneous superconducting problem. 
In detail, we set $Q_7(1,y) = n/qL$ everywhere and solve the system of equations (\ref{eqs}) 
using an ansatz for all $Q_i$ which is independent of $x$, with the boundary conditions at $y=0$. 
The solutions are shown in figure \ref{fig1}.

\begin{figure}[ptb]
\begin{subfigure}{.5\textwidth}
\centering
\includegraphics[width=0.8\linewidth]{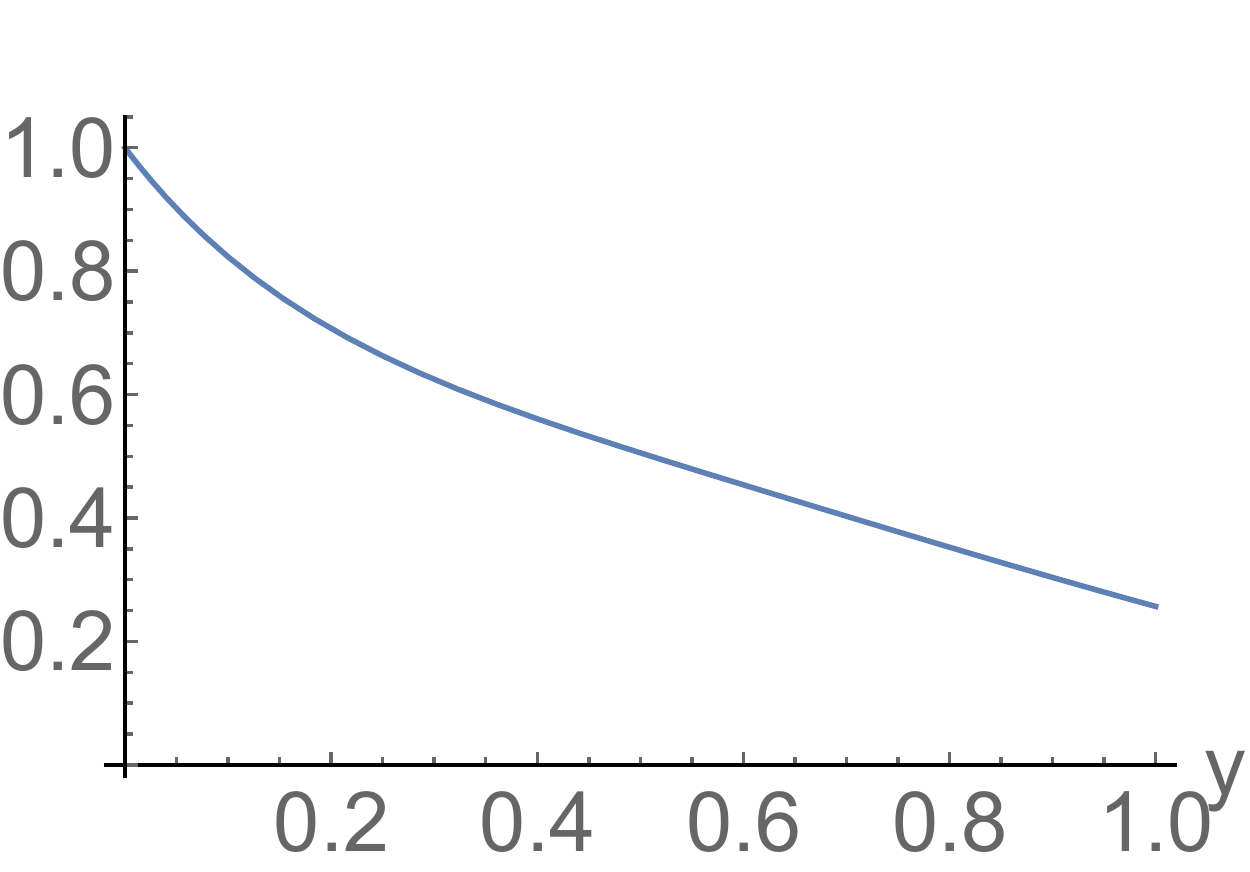}
\caption{$\tilde{Q}_1(y)$}
\end{subfigure}
\begin{subfigure}{.5\textwidth}
\centering
\includegraphics[width=0.8\linewidth]{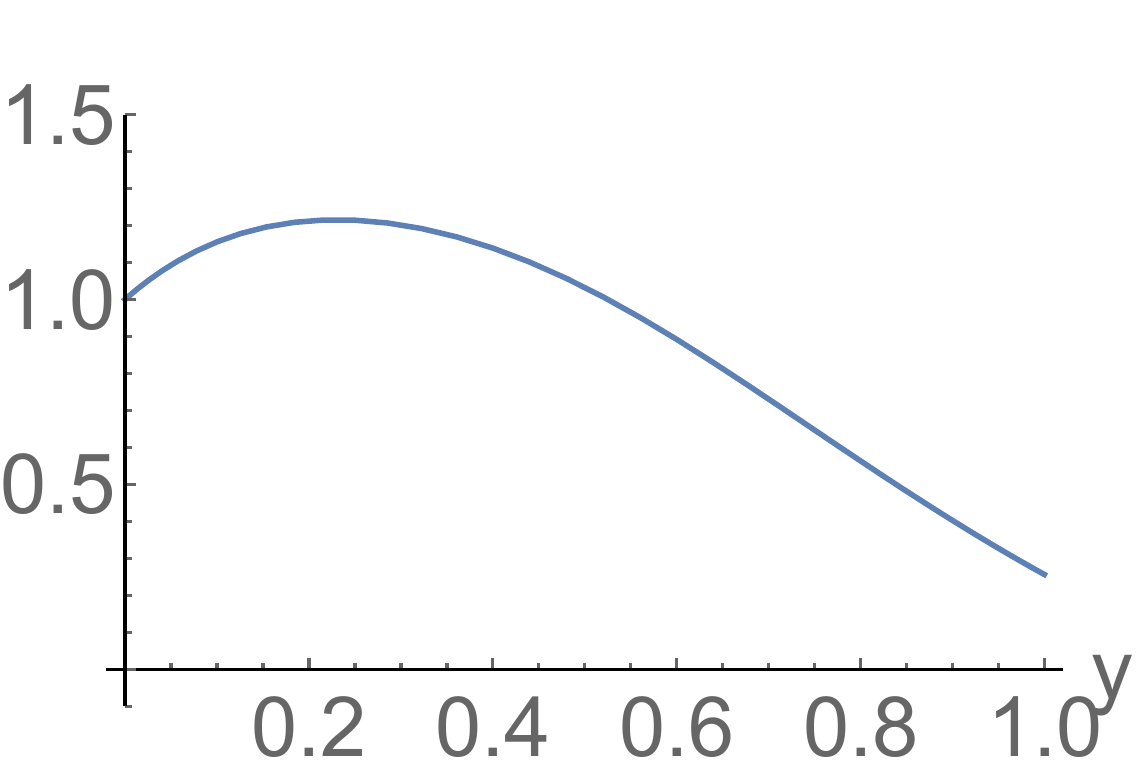}
\caption{$\tilde{Q}_2(y)$}
\end{subfigure}
\begin{subfigure}{.5\textwidth}
\centering
\includegraphics[width=0.8\linewidth]{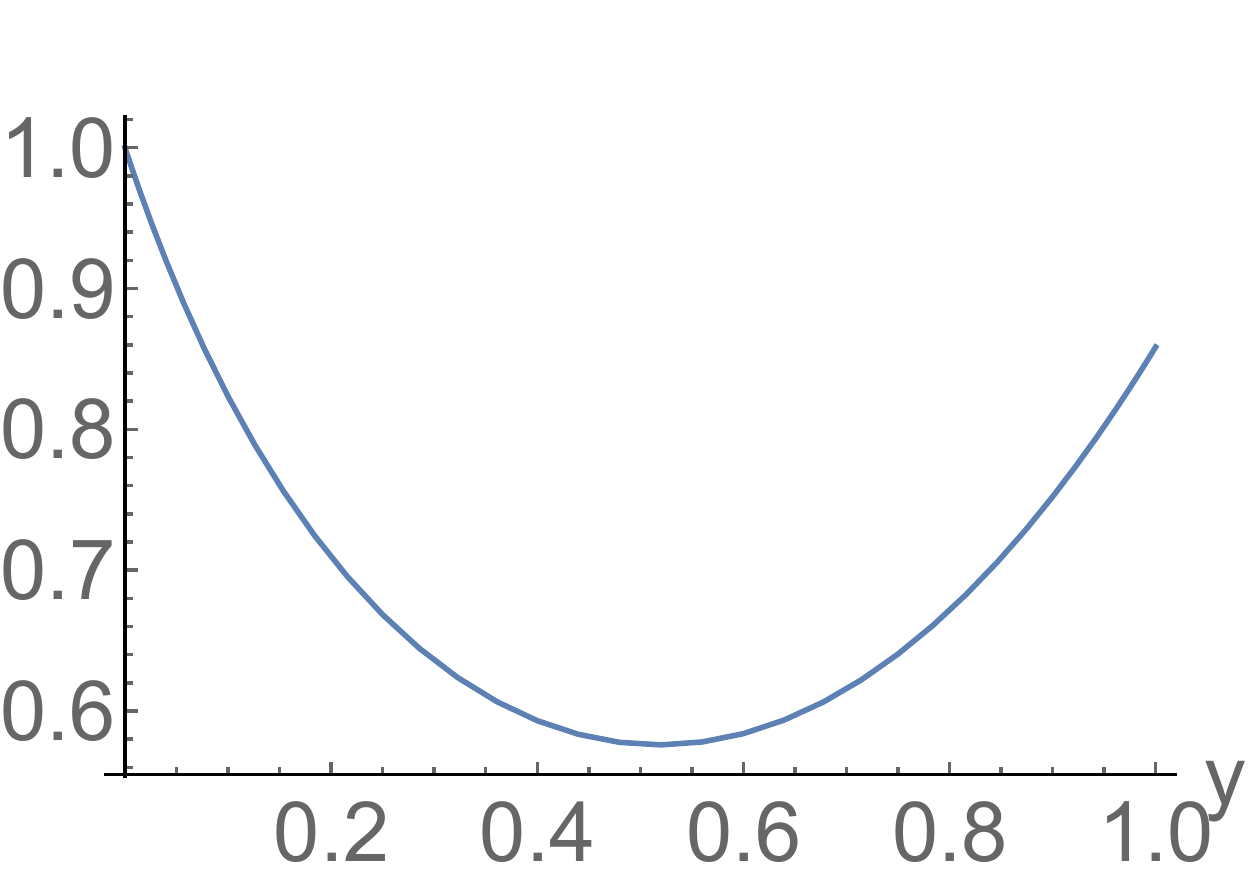}
\caption{$\tilde{Q}_3(y)$}
\end{subfigure}
\begin{subfigure}{.5\textwidth}
\centering
\includegraphics[width=0.8\linewidth]{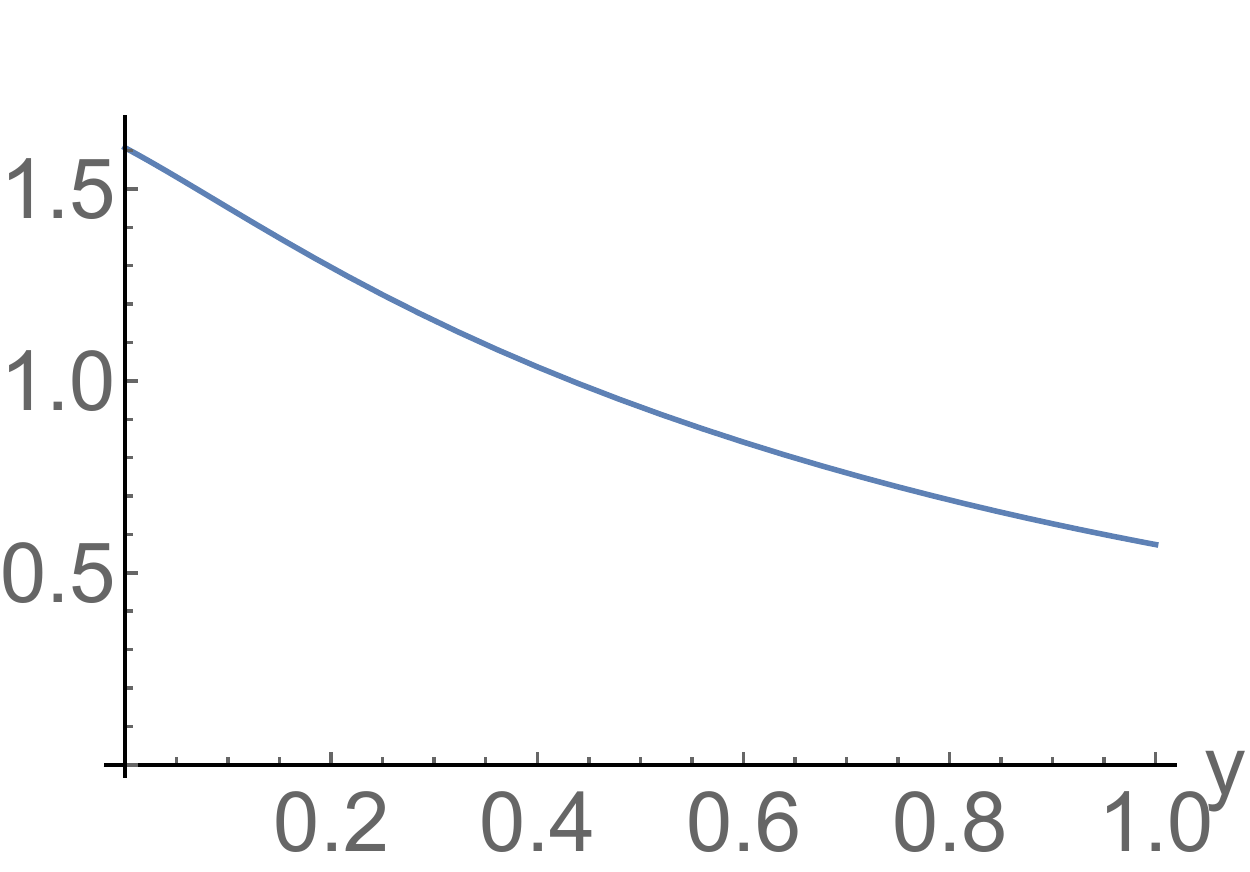}
\caption{$\tilde{Q}_4(y)$}
\end{subfigure}
\caption{Solutions of the homogeneous system representing the boundary conditions imposed on the fields at $x=1$ for $q=L=n=1$, $\kappa=-1$, $y_+=1/2$.}
\label{fig1}
\end{figure}

\item { $x=0$}

With the chosen coordinate and field definitions the boundary conditions on the fields in the vortex core where derived in the appendix of \cite{Dias:2013bwa}, they are
\bea
\partial_x Q_1(0,y)  &=&  \partial_x Q_2(0,y)=\partial_x Q_4(0,y)=\partial_x Q_5(0,y)=0  , \quad  Q_4(0,y) =  Q_5(0,y),
\nl
\partial_x Q_3(0,y) &=& 2 Q_3(0,y), \quad  \partial_x Q_6(0,y) =n Q_6(0,y) \quad \partial_x Q_7(0,y) =2 Q_7(0,y).
\eea

\item{  $y=1$} 

The reader might think that, since the system's equations are of second order, one is forced to specify conditions at the horizon as well. However this is not the case since the reparameterization used for the fields automatically enforces the correct behaviour at the horizon. The only condition that one must satisfy is that $Q_1(x,1) = Q_2(x,1)$.
\end{itemize}

\subsection{Numerical solutions}

\begin{figure}[ptb]
\begin{subfigure}{.5\textwidth}
\centering
\includegraphics[width=1.2\linewidth]{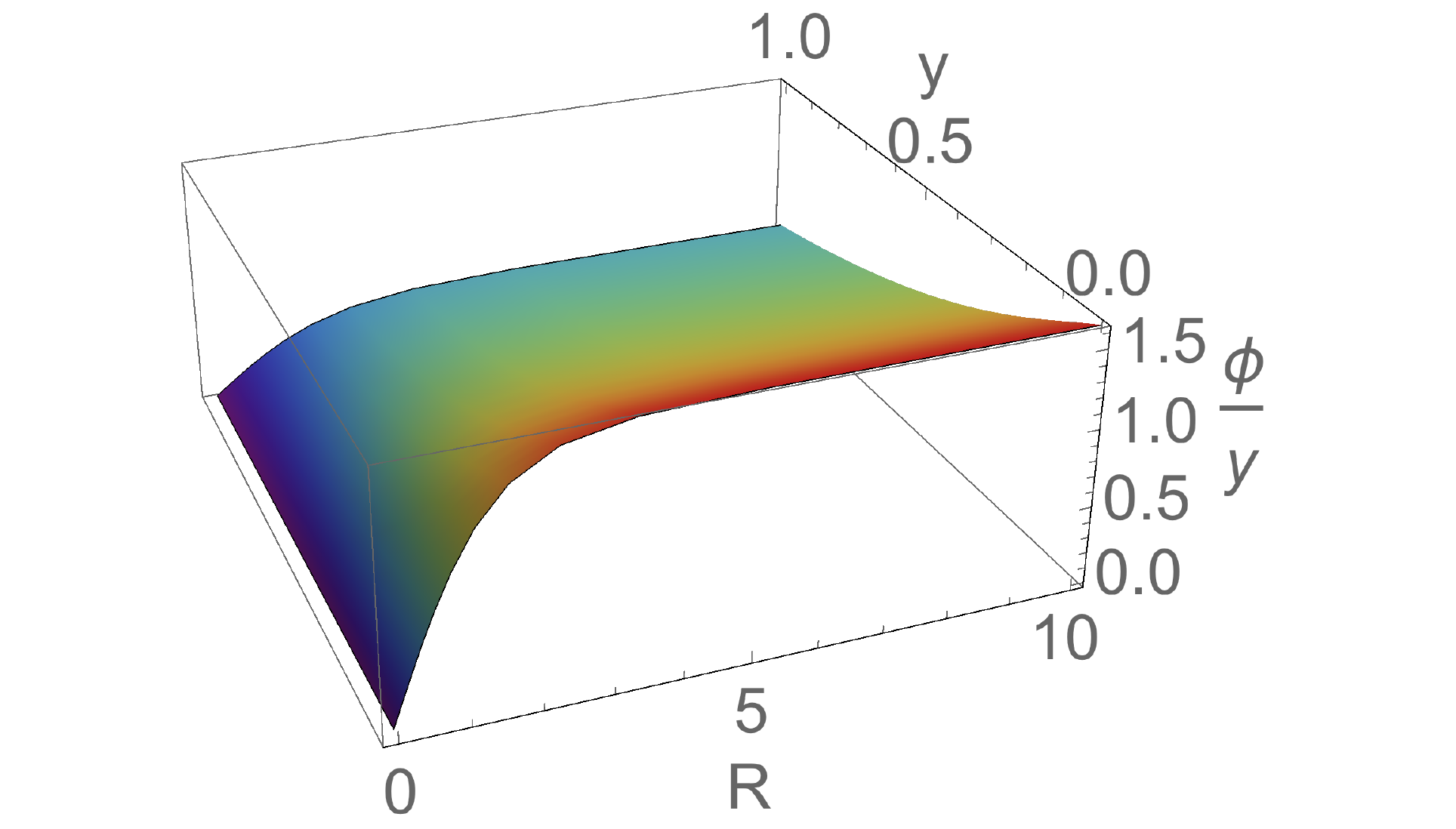}
\caption{Scalar field profile}
\end{subfigure}
\begin{subfigure}{.5\textwidth}
\centering
\includegraphics[width=\linewidth]{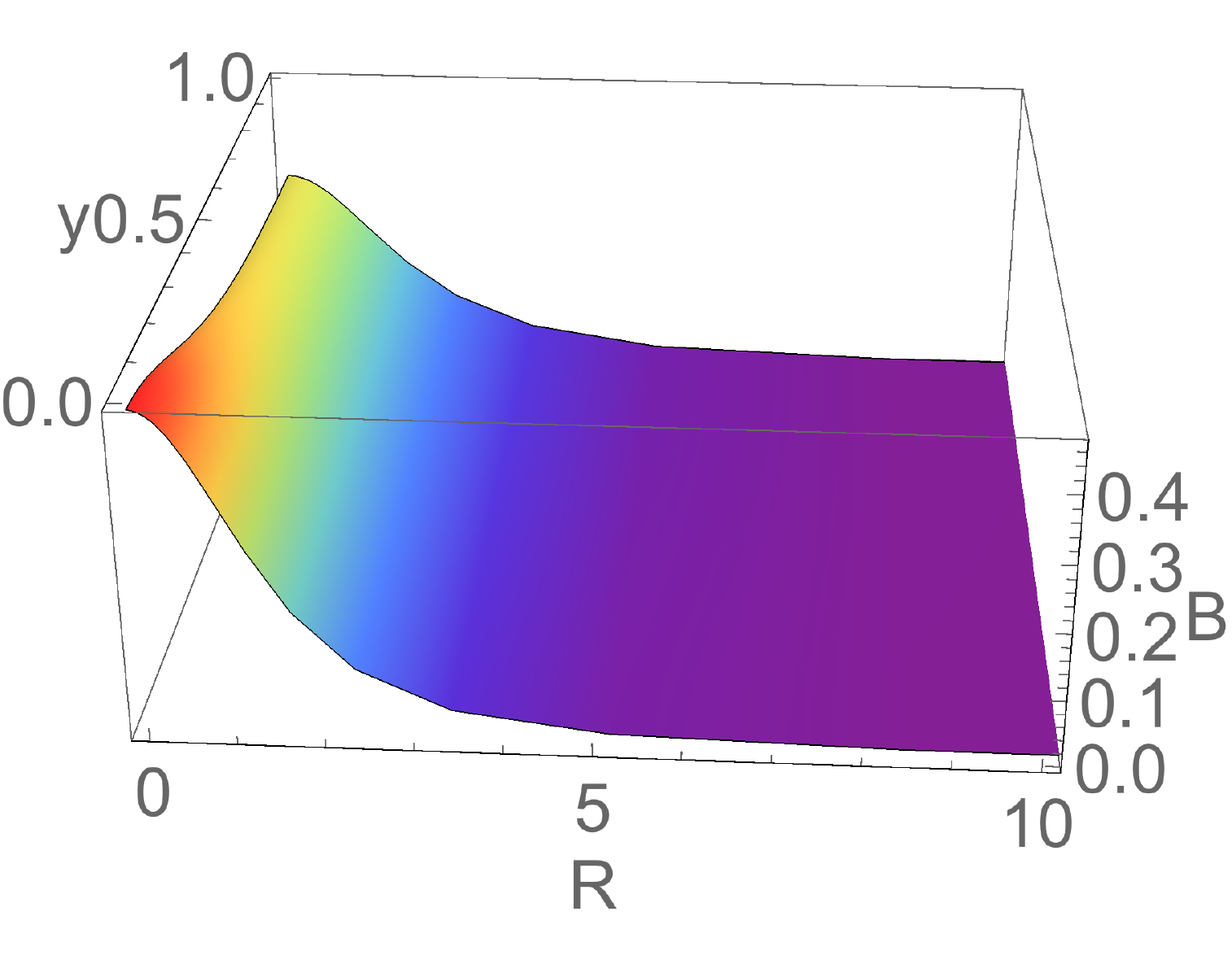}
\caption{Magnetic field profile}
\end{subfigure}
\begin{subfigure}{.5\textwidth}
\centering
\includegraphics[width=0.8\linewidth]{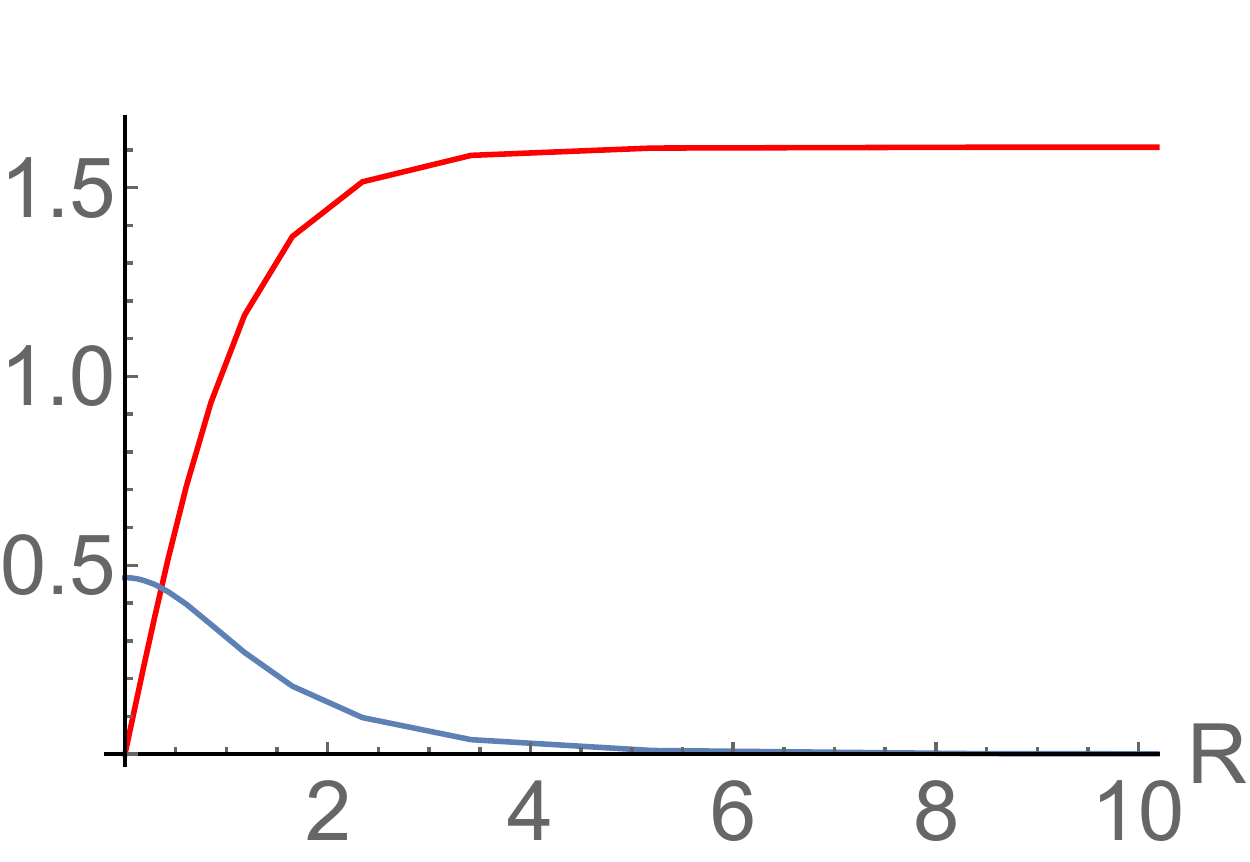}
\caption{Boundary profiles for $Q_6(x,0)$ and the magnetic field $B$ at $y=0$. Red line is scalar field, blue line is magnetic field.}
\end{subfigure}
\begin{subfigure}{.5\textwidth}
\centering
\includegraphics[width=0.8\linewidth]{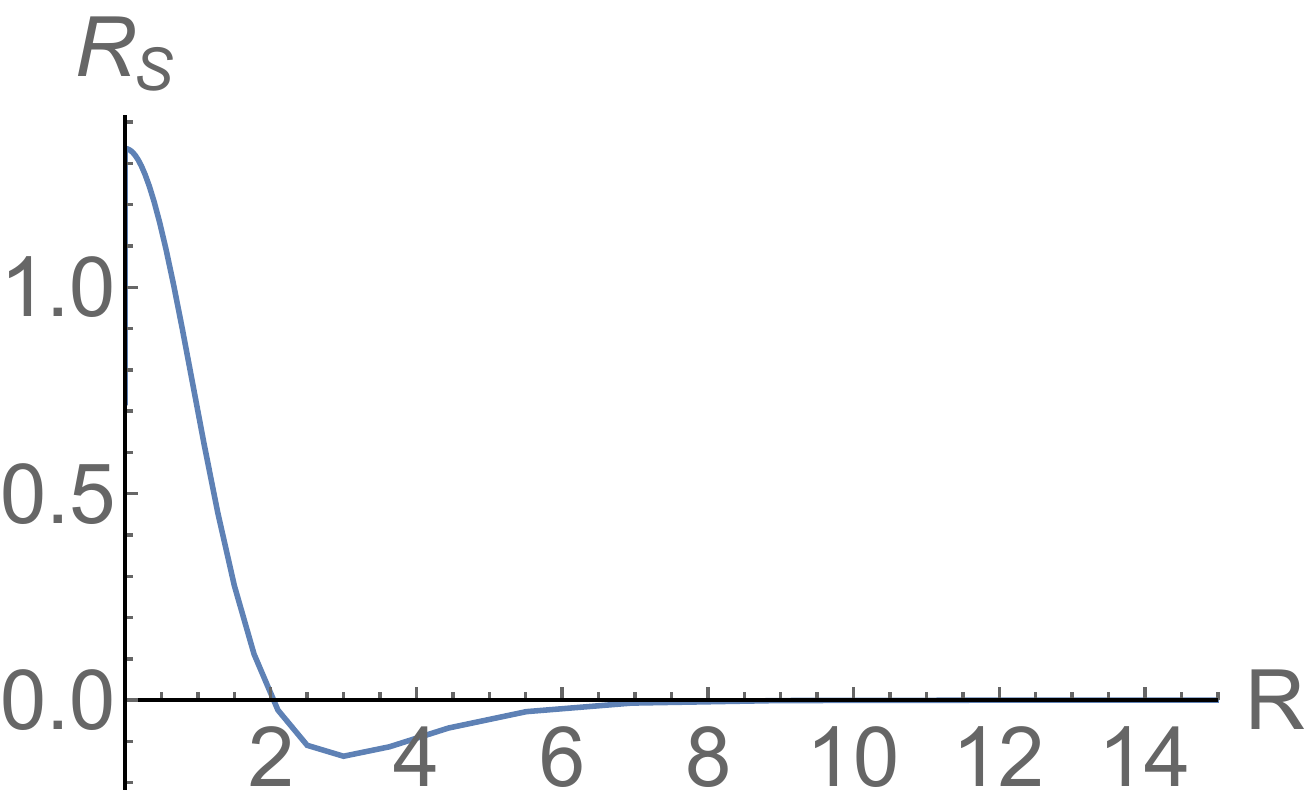}
\caption{Induced Ricci scalar $R_s$ at the horizon.}
\end{subfigure}
\caption{Solutions of the gravitating Abrikosov-like vortex at  , $q=L=n=1$, $\kappa=-1$, $y_+=1/2$, plotted as a function of the physical radial coordinate $R$.}
\label{fig2}
\end{figure}

The above system of equations (\ref{eqs}) with the set of boundary conditions defines a system we must solve. Analytic solutions cannot be found, therefore the system is solved numerically. In this paper we adopt a similar procedure to that used in \cite{Dias:2013bwa}. We use a spectral solver on a Chebyshev grid, coupled to a Newton-Rhapson linear solver. Once a suitable initial seed is given we expect exponential convergence to a solution, if indeed this exists.  Solutions are shown in Figure \ref{fig2}. They describe vortices emanating from the horizon of a planar black-hole and carrying magnetic flux. The back-reaction of these vortices deforms the horizon of the black-hole, where it forms a Reissner-Nordstrom patch (corresponding to the center of the vortex carrying magnetic flux) over an AdS-Schwarzchild solution. The deformation can be best understood by plotting the Ricci scalar $R_s$ induced at the horizon. The induced metric at the horizon is
\be
ds^2 = \frac{L^2y_+^2}{(1-x)^2}\left[\frac{Q_4(x,1)dx^2}{(1-x)^2}+x^2Q_5(x,1) d\theta^2\right],
\ee
for which we plot $R_s$ evaluated on the solution in figure \ref{fig2}-(d). For the remaining features of this solution, including its thermodynamical properties, we refer the reader to  \cite{Dias:2013bwa}.

\section{Non-abelian vortex}
\label{sect3}

In this section we will study an AdS realisation of 
a non-abelian vortex model which, in the flat spacetime
limit, reduces to the model introduced in \cite{Shifman:2012vv}.
The setup is conceptually similar to the model introduced
by Witten \cite{Witten:1984eb} for cosmic strings, and can be generalized to monopoles \cite{Shifman:2015ama} 
and skyrmions \cite{Canfora:2016spb}. The main difference between the
flat and the AdS case is that in the former case the non-abelian symmetry responsible
for the orientational zero modes is global, while in the latter it is a gauged
symmetry in the AdS bulk (which, by holographic dictionary, corresponds to
a global symmetry in the boundary dual theory).

A very similar model for the non-abelian vortex  in AdS was previously studied
in \cite{Tallarita:2015mca},  neglecting the vortex gravitational back-reaction. 
The action that we will consider is as follows:
\be
S_T = S_{ANO}+S_\chi \, ,
\ee
where $S_{ANO}$ is the action shown in equation (\ref{sano}) and 
\be
S_\chi = \frac{1}{16\pi G_N}\int d^4x\sqrt{-g}\left[-\frac{1}{2 g_2^2} \Tr  \, \tilde{F}_{\mu\nu}\tilde{F}^{\mu\nu}-\frac{1}{2}\tilde{D}_\mu \chi^i \tilde{D}^\mu \chi_i - \tilde{V}(|\phi|^2,\chi^2)\right].
\label{esse-chi}
\ee
Here, $\chi$ is a real scalar field in the adjoint representation of $SU(2)$
 such that $\chi = \chi^i \frac{\sigma_i}{2}$ with $\sigma_i$ the standard Pauli matrices, and $\chi^2 = \chi^i\chi_i$. 
Moreover, $\tilde{F}_{\mu\nu}=\tilde{F}_{\mu\nu}^i \frac{\sigma_i}{2}$ is 
the non-abelian field strength of the $SU(2)$ gauge field $\tilde{A}_\mu=\tilde{A}_\mu^i \frac{\sigma_i}{2}$, i.e.
\beq
\tilde{F}_{\mu\nu}^a=\p_\mu  \tilde{A}^a_\nu -  \p_\nu  \tilde{A}^a_\mu+\epsilon^{abc} \tilde{A}^a_\mu \tilde{A}^b_\nu \, ,
\eeq
The symbol $\tilde{D}_\mu$ denotes in general the combination
of the gravitational and $SU(2)$ gauge covariant derivatives,
while $\hD_\mu$ denotes a partial derivative with the inclusion of only the $SU(2)$ covariant
term; for scalars the two definitions coincide: 
\be
\tilde{D}_\mu \chi_a = \hD_\mu \chi_a = \p_\mu \chi_a  +  \epsilon_{abc} \tilde{A}_\mu^b\chi^c,
\ee
In contrast to \cite{Tallarita:2015mca}, no additional $U(1)$ sector is needed 
as we follow the guideline of adding no chemical potential in the dual field theory. 
Furthermore, note that the $\chi$ sector is neutral with respect to the original local 
$U(1)$ symmetry under which $\phi$ is charged.  

The extra term in the potential is chosen to be of the form:
\be
\label{potechi}
\tilde{V}(|\phi|^2,\chi^2) = -\frac{1}{L^2} |\chi|^2
 +\frac{\gamma}{L^2} |\phi|^2 \chi^2 + \frac{\gamma    \beta}{L^2}  \chi^4  \, .
\ee
The purpose of the mixed $|\phi|^2 \chi^2 $ term is to make
 the condensation of $\chi$ energetically favorable
in the region where the condensate of $\phi$ is zero,
which corresponds to the center of the vortex.
In this way the field $\chi$ condenses
just in the center of the vortex, where the non-abelian
orientational model will be localized.

The original equation of motion for the gauge sector, equation (\ref{eqs}), is unchanged, 
while the equations for the scalar fields $\phi$ and $\chi$ become,
\be
g^{\mu \nu}D_\mu D_\nu \phi-\left[V'(|\phi|^2)+\frac{\gamma\chi^2}{4 L^2} \right]\phi=0 \, ,
\ee
\be\label{chieq}
g^{\mu \nu} \tilde{D}_\mu \tilde{D}_\nu \chi^i
-\frac{(-2 + 2\gamma |\phi|^2+4 \gamma \beta \chi^2)}{L^2} \chi^i=0 \, .
\ee
The coupled potentials for $\phi$ and $\chi$ lead to an interesting range of vacua, these are
\bea
&(I)& \qquad \phi=0 \, , \qquad \chi=0 \, ,
\nl
&(II)& \qquad  \phi=0, \quad \chi^2 = \frac{1}{2\beta \gamma},
\nl
&(III)& \qquad |\phi|^2=1, \quad \chi^2 = 0,
\nl
&(IV)& \qquad |\phi|^2=\frac{4 \beta-1}{4 \beta -\gamma} , \quad 
\chi^2 = \frac{2-2 \gamma}{\gamma (4 \beta-\gamma)}.
\eea
The asymptotic AdS vacuum that we choose corresponds to the
first of these vacua.
In this vacuum, the quadratic part of the potential is chosen in 
such a way that
\beq
m^2_\chi =m^2_\phi = -2/L^2 \,.
\eeq
The boundary expansion of the field $\phi$ in the AdS vacuum (\ref{empty-ads}) is unchanged (see eq.~(\ref{bou-phi})),
while the boundary expansion of the adjoint field $\chi_i$ is:
\be
\chi_i = \alpha_i z + \beta_i z^2+\dots \, .
\label{chi-asint}
\ee
We will consider a Robin condition of the form
 $\beta_i = \eta \alpha_i$, which is dual in the boundary theory potential
 to a double trace deformation of the form:
 \be
 \Delta \mathcal{V}= \eta \, (\mathcal{O}_1^2+\mathcal{O}_2^2+\mathcal{O}_3^2) \, ,
\ee
where $\mathcal{O}_i$ are the dual operators to the $\chi_i$ bulk fields. \newline

The equations of motion for the $SU(2)$ gauge field are:
\beq
\frac{1}{4 g_2^2}
\frac{1}{  \sqrt{-g}} \hD_\mu (  \sqrt{-g} g^{\mu \a} g^{\nu \b} \tilde{F}^a_{\a \b}) 
+\epsilon^{abc} \chi^b \hD_\mu \chi^c = 0 \, .
\label{gau-su2}
\eeq
For the static vortex solution, we will consider a configuration where the
gauge field is set to zero and the only non-vanishing component
of the non-abelian scalar is $\chi^3$, so eq. (\ref{gau-su2}) is automatically satisfied.
This equation will be later useful in order to study zero modes localized on the vortex.

 Since we are interested in including the back-reaction, 
 the gravitational sector is also modified and becomes
\be
G_{\mu \nu}^T=G_{\mu \nu}-G_{\mu \nu}^\chi=0,
\ee
where $G_{\mu \nu}$ is given by equation (\ref{eqs}) and
\be
G_{\mu \nu}^\chi=\left[(\tilde{D}_\mu \chi)(\tilde{D}_\nu \chi)+(\tilde{D}_\nu \chi)(\tilde{D}_\mu \chi)
+g_{\mu \nu}\tilde{V}(|\phi|^2,\chi^2)
+ \Tr \left(\tilde{F}_\mu^\rho \tilde{F}_{\nu \rho}-\frac{g_{\mu \nu}}{4}\tilde{F}^{\rho \sigma }
\tilde{F}_{\rho \sigma }\right)\right].
\ee
Therefore we look for solutions of this new system, with an extra $\chi$ field.
For the vortex solution, we will use the ansatz $\chi^1=\chi^2=0$
and we will parametrize the profile of the
field $\chi^3$ with the function $Q_8$, i.e.
\be
\chi^3 = y\; Q_8(x,y) \, .
\ee

\subsection{Boundary conditions of the $\chi$ field}

The boundary conditions for all the matter fields and gravitational sector were discussed previously, 
we now discuss those for the additional field $\chi$.

\begin{itemize}

\item {y=0}

To discuss the boundary conditions at the asymptotic boundary for $\chi$ we must first analyse the equation it satisfies there. 
First of all, in order for the $\chi$ field to have an asymptotic expansion of the form
\be
Q_8(x,0) = \hat{\chi}_1 (x) + y \, \hat{\chi}_2 (x) + \dots \, .
\ee
Taking a series expansion of the equation of motion at $y =0$ we get the following condition
\be\label{condition}
Q_1'(x,0) +3 Q_2'(x,0)+Q_4'(x,0)+Q_5'(x,0)=0,
\ee
where $'$ denotes differentiation w.r.t $y$. This equation is obtained at order $y^1$ unless $Q_8(x,0)=0$. 
This equation is a non-trivial condition on the metric fields, which has to be satisfied if we want $Q_8$ to appear.  

Regarding the actual condition on $Q_8$: in \cite{Tallarita:2015mca}
 this field was made to condense by adding a separate chemical potential for it. 
 This in turn implied an additional $U(1)$ gauge symmetry in the bulk, 
 under which only $\chi$ was charged. However, as we learnt from \cite{Faulkner:2010gj} and
  are using throughout this paper, the addition of the chemical potential can be replaced 
  by a suitable relevant operator in the boundary theory. 
  Therefore for the boundary variable $Q_8$, we also use a Robin like boundary condition of the form
\be
\partial_y Q_8= \frac{ \kappa_2 }{y_+}Q_8.
\ee
In turn, this Robin condition has a holographic interpretation
in term of the double trace deformation parameter $\eta$;
the precise identification depends on  the  $\tilde{\alpha}$
parameter in the DeTurck reference metric and will be discussed
in the next section,  see eq. (\ref{kap}).

\item{x=0}

At the vortex core, we require that the $\chi$ field satisfies Neumann conditions. 
In turn this amounts to picking Neumann conditions for $Q_8$, therefore
\be
\partial_x Q_8(0,y)=0.
\ee

\item{x=1}

Infinitely far away from the vortex core, the $\chi$ field should approach its asymptotic vacuum value. 
This is given by the precise choice of parameters in the potential, and each choice corresponds to a specific vacuum. 
Therefore, we will consider general Neumann conditions of the type
\be
\partial_x Q_8(1,y)=0,
\ee
and let the parameters pick the right vacuum. In particular note that for the vacuum we are interested in, for which $Q_8 = 0$, the gravitational solutions shown in Figure \ref{fig1} are still valid.

\end{itemize}

\subsection{Numerical solutions}

\begin{figure}[ptb]
\begin{subfigure}{.5\textwidth}
\centering
\includegraphics[width=1.1\linewidth]{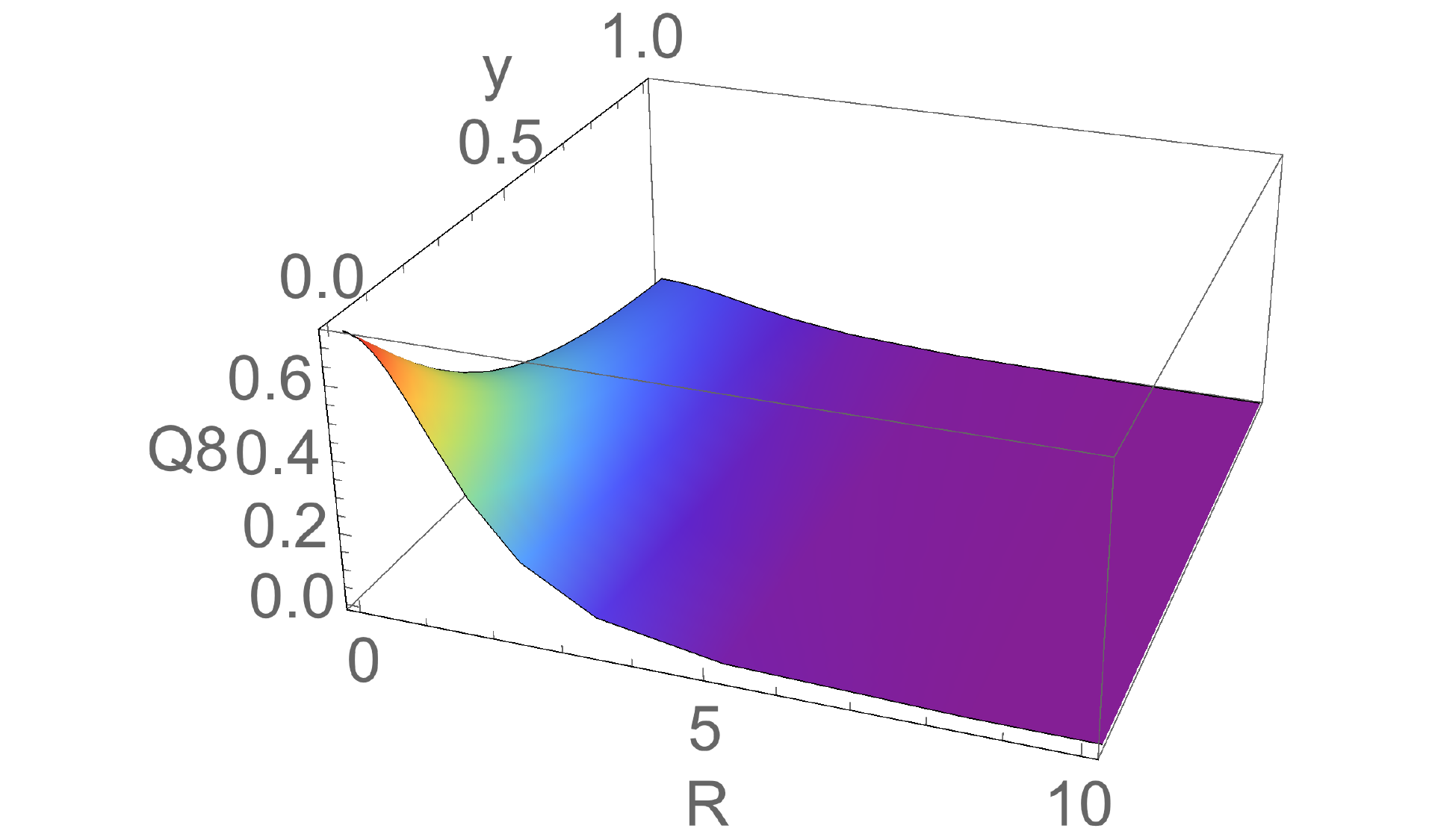}
\caption{$Q_8(x,y)$ field profile}
\end{subfigure}
\begin{subfigure}{.5\textwidth}
\centering
\includegraphics[width=0.8\linewidth]{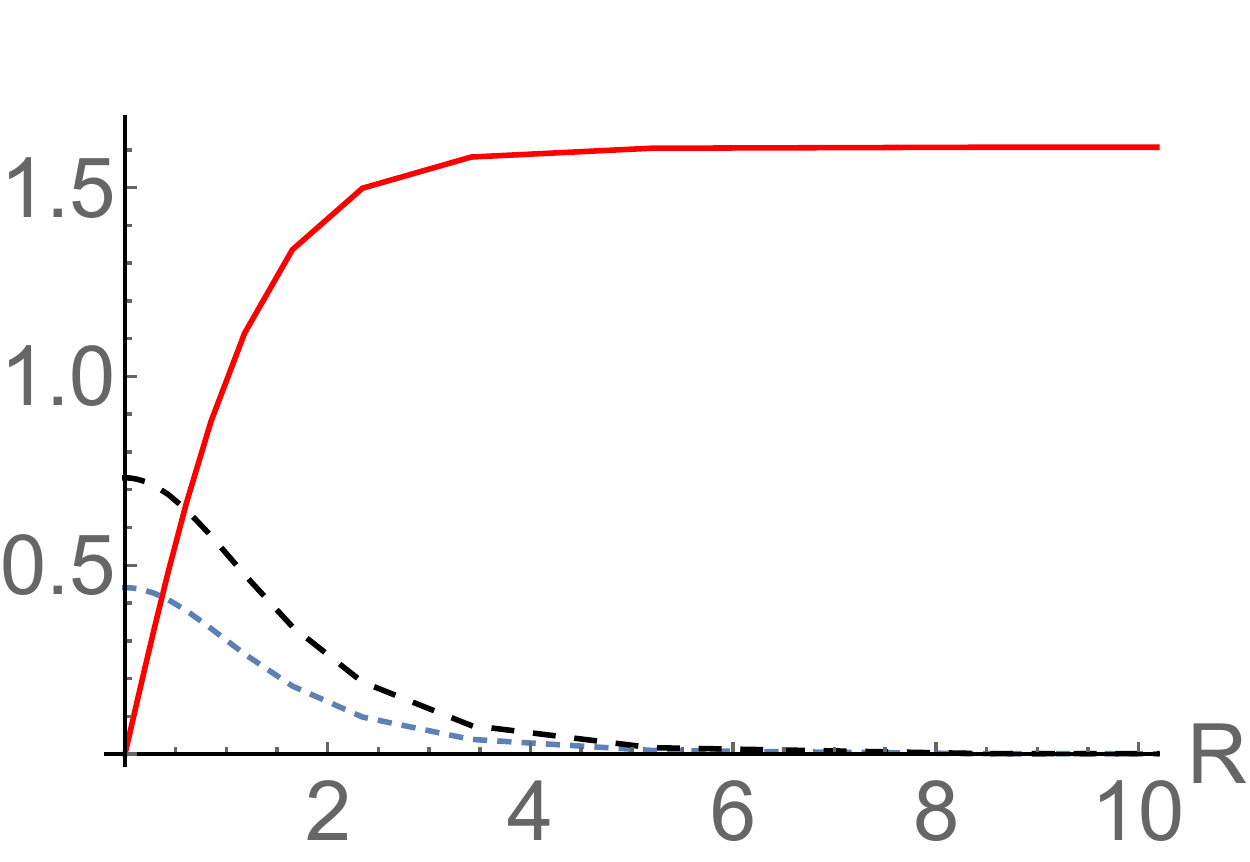}
\caption{Boundary profiles for magnetic field (small dashing), scalar condensate (solid line) and $\chi$ condensate (medium dashing).}
\end{subfigure}
\caption{Solutions of the gravitating non-abelian vortex at  , $q=L=n=1$, 
$\kappa=-1$, $\tilde{\kappa} = 2\kappa$, $y_+=1/2$, $\gamma = 3$, $\beta = 1$, 
plotted as a function of the physical radial coordinate $R$.}
\label{fig3}
\end{figure}

Therefore we have a full set of equations with boundary conditions which we must solve. 
These have to be solved numerically and we use a similar procedure to that outlined above. 
A solution is shown in figure \ref{fig3}. 
The remaining fields vary little from the solutions presented in figure \ref{fig2} and therefore we choose not to present them here. The boundary profiles presented above are, as expected, similar to the flat space solutions
 obtained for example in  \cite{Shifman:2014oqa} and represent a non-abelian vortex in the dual theory . \newline

For all the solutions presented in this paper we have verified numerically that the norm of the DeTurck vector  $\xi^a\xi_a =0$ and that the condition outlined in eq.(\ref{condition}) are satisfied to accuracy $\mathcal{O}(10^{-7})$.

\section{Free energy}
\label{sect4}

This section is devoted to the study of thermodynamic quantities in the holographic theory.
 In particular, we are interested in showing that the solutions in which
  the additional field $Q_8$ condenses are thermodynamically preferred over the solutions in which it doesn't. 
   In order to compute the free energy we need first to find the holographic
energy momentum tensor of the boundary field theory. 
In order to perform the necessary holographic renormalization procedure \cite{Balasubramanian:1999re},
it is convenient to use Fefferman-Graham (FG) coordinates:
\beq
ds_{FG}^2=\frac{L^2}{z^2} dz^2 + \gamma_{M N} dw^M dw^M   \, , 
\eeq
where the capital latin letter denote the boundary coordinates $w^M=(t, \tx,  \theta)$
and $y \approx y_+ z$.   Near the boundary
\beq
\gamma_{M N} \approx \frac{h_{MN}}{z^2} \, , \qquad
h_{M N} dw^M dw^N = -dt^2 +\frac{d \tx^2}{(1-\tx)^4} + \frac{\tx^2}{(1-\tx)^2} d \theta^2 \, ,
\eeq
which is the flat space metric.
Since we want to extract holographic quantities at the boundary, 
it is sufficient to perform this coordinate transformation 
as a series expansion nearby $y=0$. A similar procedure was performed in \cite{Dias:2013bwa}.

It is useful to write the functions $Q_i$ as a series expansion
in powers of $y$ around $y=0$:
\beq
Q_i= \sum_{k=0}^\infty Q_i^{(k)} y^k \, .
\label{ee}
\eeq
The coefficients $Q_i^{(k)}$ of the expansion can be determined recursively 
expanding the equations of motion as a function of $y$. 
We find that these coefficients can all be obtained as functions of:
\beq
Q_3^{(2)} \, , \qquad
Q_4^{(3)} \, , \qquad Q_6^{(0)}  \, , \qquad  Q_7^{(0)}  \, , \qquad  Q_8^{(0)}
\eeq
 (we have checked this claim up expanding eqs up to 6th order in $y$). 
 These series expansion coefficients depend explicitly on the
 value of $\ta$ in the DeTurck reference metric (\ref{alphatilde}).

The FG coordinates $(z,\tx)$ are related to $(y,x)$ by the following expansion:
\beq
y=y_+ z + \sum_{i=2}^\infty a_i(\tx) z^i \, , 
\qquad x= \tx + \sum_{i=1}^\infty b_i(\tx) z^i \, .
\label{ee-ab}
\eeq
The first coefficients $(a_i,b_i)$, found by expanding 
the metric close to the boundary, as well as
using equations of motions for the $Q_i$ coefficients, are given in appendix A. 
From the change of variables, we find the match between $\kappa$ and $\kappa_1$ and $\eta$ and $\kappa_2$:
\beq
\kappa_1=	\kappa -\frac{5 \ta  y_+}{16}  \, , \qquad
\kappa_2=	\eta -\frac{5 \ta  y_+}{16}  \, ,
\label{kap}
\eeq
 
 As explained in \cite{Dias:2013bwa}, for generic values of $\ta$
 in the DeTurck reference metric (\ref{alphatilde}),
  the series expansion in eqs. (\ref{ee},\ref{ee-ab})
should be generalized in order to include $\log y$ terms.
 In order to avoid these logarithms, which make the holographic renormalization
 procedure much more complicated, 
 one can choose to the following value of $\ta$:
 \beq
\ta=4 \kappa_1/y_+ \, , 
\label{alfetta}
\eeq 
which, combined with eq. (\ref{kap}), gives $\kappa_1 =\frac{4}{9} \kappa$.
Ref. \cite{Dias:2013bwa} indeed checked that, for this value of $\ta$,
there are no $y^k \log y$  terms up to tenth order in $k$.
A good test of the consistency of this procedure comes from 
the first law of thermodynamics, as we shall see later. 

In our example we have two Robin conditions 
for each of the fields $\phi$, $\chi^i$.
In order to avoid logarithms in the series expansion in $y$, we specialize to
 \be
 \label{rest}
 \kappa_1 = \kappa_2 \, ,
 \ee
and we use the $\ta$ given by eq. (\ref{alfetta}).

We can then use the results of \cite{Balasubramanian:1999re} to extract the energy momentum tensor:
\beq
T_{MN}=\frac{1}{8 \pi G_N L^2} \,  \lim_{z \rightarrow 0}
\frac{L}{z} \le K_{MN} -\gamma_{MN} K -\frac{2}{L} \gamma_{MN} 
-\frac{\gamma_{MN}}{L}  \le \phi^2 + \frac{\chi^2}{2} \ri  \ri \, ,
\eeq
where $K_{MN}, K$ are the extrinsic curvature tensor and scalar calculated with 
an inward unit normal vector to the constant
$z$ surfaces nearby the boundary. For later convenience, let us introduce the operator VEVS:
\beq 
 |\langle \mathcal{O}_1 \rangle |= y_+ Q_6^{(0)} x^n \, , \qquad
  |\langle \mathcal{O}_2 \rangle |= y_+ Q_8^{(0)}  \, .
\eeq
In presence of double-trace deformation, $T_{MN}$ is  not covariantly conserved:
\beq
D^M T_{MN} = \frac{55 \alpha  y_+ -112 \kappa_1}{384 \pi  G}
 D_N |\langle \mathcal{O}_1  \rangle|^2
+ \frac{55 \alpha  y_+ -120 \kappa_1+8 \kappa_2}{768 \pi  G}
 D_N |\langle \mathcal{O}_2  \rangle|^2 \, ,
\eeq
 where $D_N$ denotes the boundary covariant derivative.
Moreover, the trace anomaly reads:
\beq
h^{MN} T_{MN}=\frac{\kappa}{4 \pi G} |\langle \mathcal{O}_1 \rangle |^2
+\frac{5 \kappa + 9 \kappa_2}{72 \pi G} |\langle \mathcal{O}_2 \rangle |^2 \, .
\eeq
Therefore, one introduces a modified $\tilde{T}_{MN}$ which 
is conserved \cite{Papadimitriou:2007sj,Caldarelli:2016nni}:
\beq
\tilde{T}_{MN}= T_{MN} - h_{MN}  
\le \frac{55 \alpha  y_+ -112 \kappa_1}{384 \pi  G} |\langle \mathcal{O}_1 \rangle |^2 
+  \frac{55 \alpha  y_+ -120 \kappa_1+8 \kappa_2}{768 \pi  G}   | \langle \mathcal{O}_2 \rangle |^2 \ri \,.
\eeq

The energy density can finally be extracted as
\be
E = - \int d^2x \sqrt{\eta}\; \tilde{T}_{MN} (\partial_t)^M t^N,
\ee
where $\eta_{MN}$ is the induced metric on te constant $t$ surface 
with unit normal $t^N$. The calculated expression for the energy density is:
\beq
E= \frac{y_+^2}{G} \int \frac{\tx d \tx}{(1-\tx)^3} \le
-\frac{3}{8} y_+ Q_1^{(3)} 
- \frac{160 \kappa_1+17 \alpha y_+ }{256} \tx^{2 n} {Q_6^{(0)}}^2
- \frac{160 \kappa_2+17 \alpha y_+ }{512}  {Q_8^{(0)}}^2 
\ri \, ,
\eeq
from which the energy of the BH solution without vortex
should be subtracted (the ``vacuum"). Note that there is an expected symmetry
  between the two scalar fields $x^n Q_6^{(0)} $ and $ \frac{Q_8^{(0)}}{\sqrt{2}} $, which is why 
  we chose to present these results in terms of general $\alpha$ first. 
  Using eqs. (\ref{alfetta},\ref{rest}) , we find the expression of the regulated energy density difference as
\bea
\Delta E &=&  \frac{y_+^2}{G} \int_0^1 \frac{\tx d \tx}{(1-\tx)^3} 
\le -\frac{3}{8} y_+ (Q_1^{(3)}(\tx)-Q_1^{(3)}(1))  \right. \nl
&& \left. - \frac{57\kappa_1 }{64} (\tx^{2 n} {Q_6^{(0)}(\tx)}^2-(Q_6^{(0)}(1))^2)
- \frac{57\kappa_1}{128}  {Q_8^{(0)}(\tx)}^2 
\ri \, .
\eea

The entropy difference between our solution and the ``vacuum" is \cite{Dias:2013bwa}
\beq
\Delta S= \frac{\pi\; y_+^2}{2} \int_0^1 \frac{\tx d \tx}{(1-\tx)^3} \le
\sqrt{Q_4(x,1)Q_5(x,1)}-\tilde{Q}_4(x,1)
\ri \, .
\eeq

With these thermodynamic variables we can finally compare, in the canonical ensemble, 
the free energy of the solution with and without $\chi$. Therefore, we look at 
\be
\Delta F= \Delta E - T \Delta S,
\ee
and in particular at $\Delta F_{diff}  =\Delta F_{\chi\neq 0} - \Delta F_{\chi =0}$, 
where a negative result would indicate that the solutions with $Q_8$ are preferred.

\begin{figure}[ptb]
\begin{subfigure}{.5\textwidth}
\centering
\includegraphics[width=0.9\linewidth]{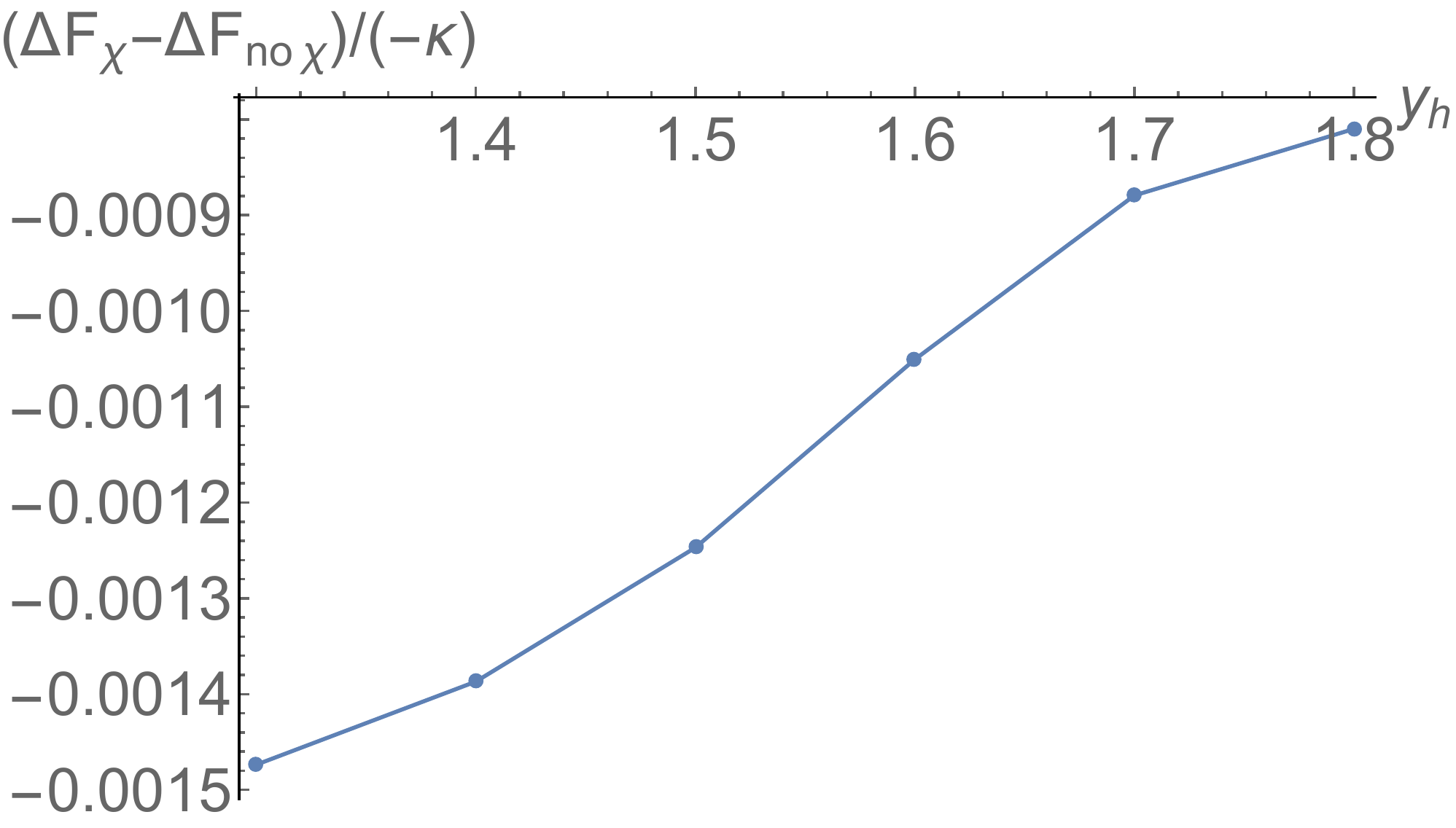}
\caption{$n=1$}
\end{subfigure}
\begin{subfigure}{.5\textwidth}
\centering
\includegraphics[width=0.9\linewidth]{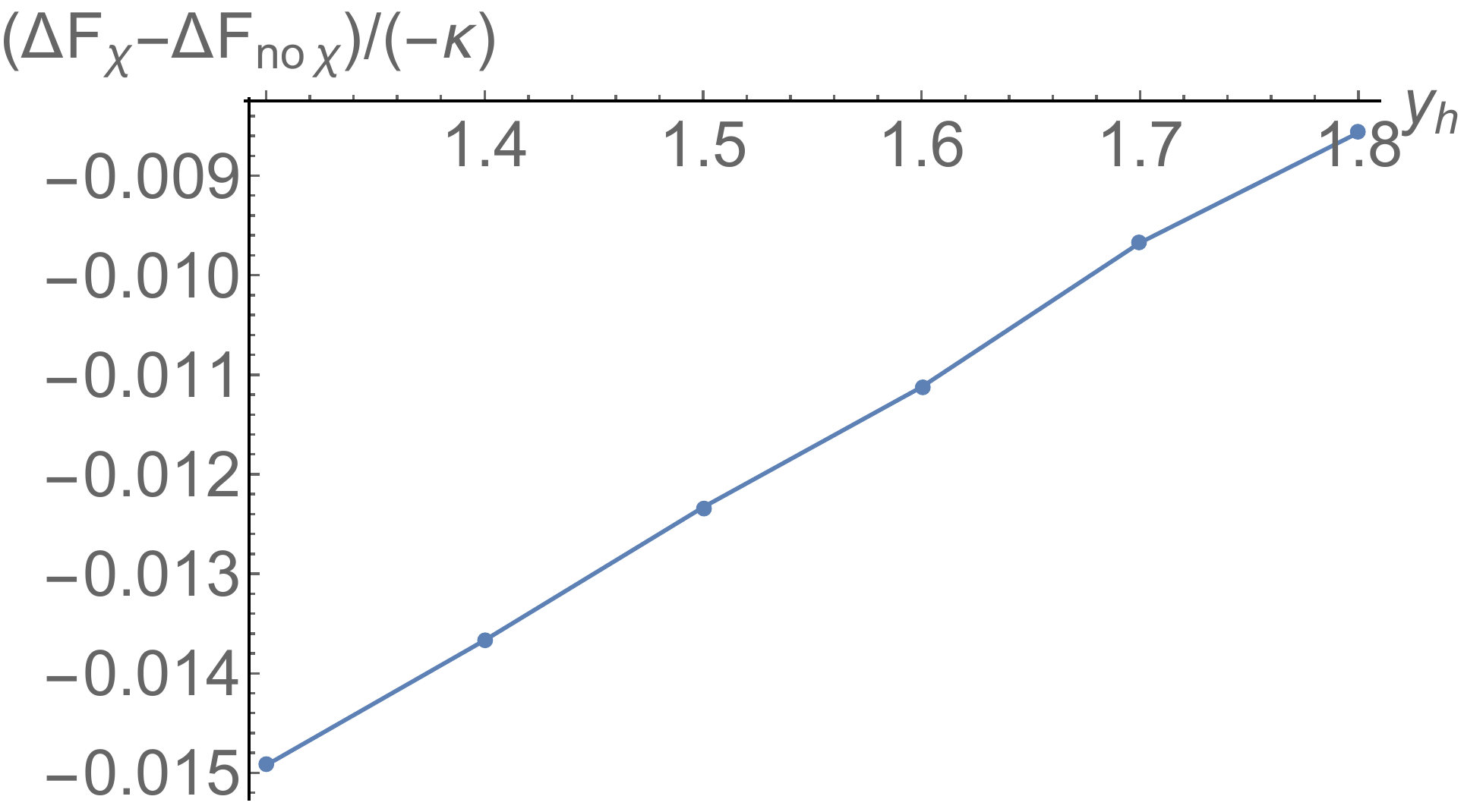}
\caption{$n=2$}
\end{subfigure}
\caption{Difference in free energy between solutions with the $\chi$ condensate and those without, as a function of $y_h$ (or temperature). A negative result indicates the phase with $\chi$ is preferred. 
The parameters $q L=1$ and $\kappa_2= \kappa_1$, $\kappa_1=-1$, $\gamma =1.1 $, $\beta=0.5$ are used.}
\label{fig-free-energy}
\end{figure}

\begin{figure}[ptb]
\begin{subfigure}{.5\textwidth}
\centering
\includegraphics[width=\linewidth]{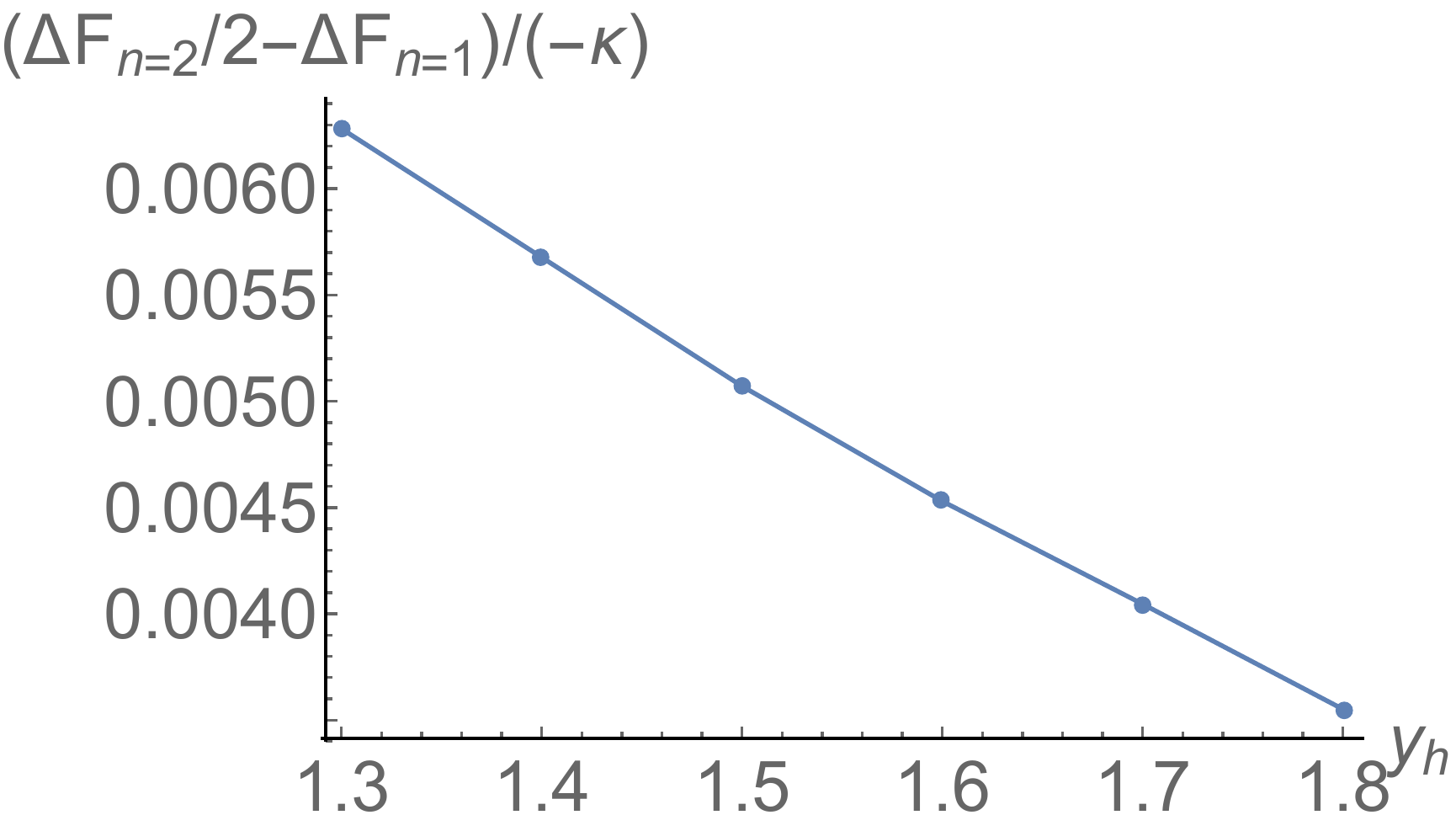}
\caption{$\chi=0$}
\end{subfigure}
\begin{subfigure}{.5\textwidth}
\centering
\includegraphics[width=\linewidth]{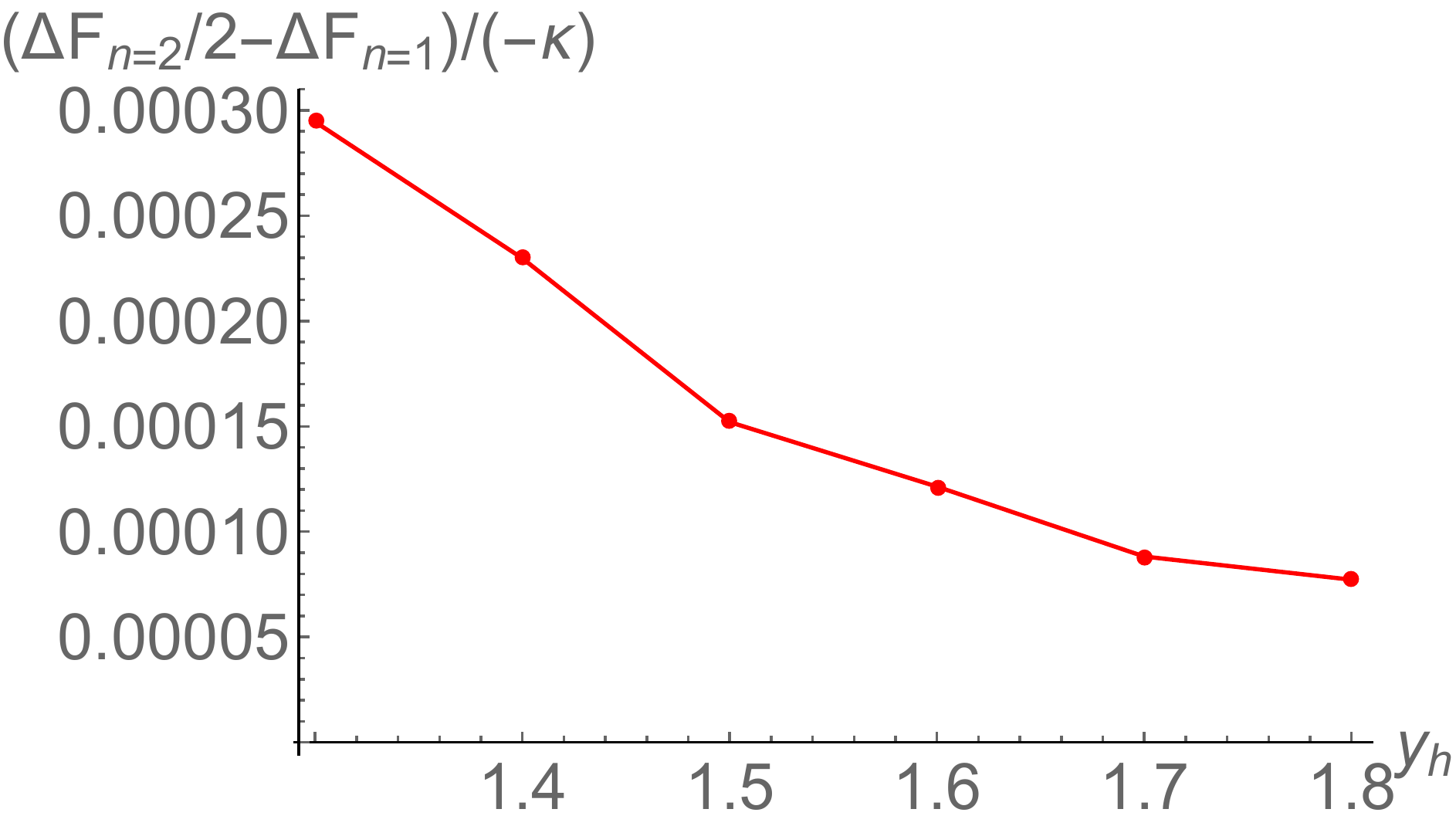}
\caption{$\chi\neq0$}
\end{subfigure}
\caption{Check of vortex type with same parameters as above, $qL =1$. For these parameters these solutions are of type II.}
\label{fig-free-energy-2}
\end{figure}

\begin{figure}[ptb]
\begin{subfigure}{.5\textwidth}
\centering
\includegraphics[width=\linewidth]{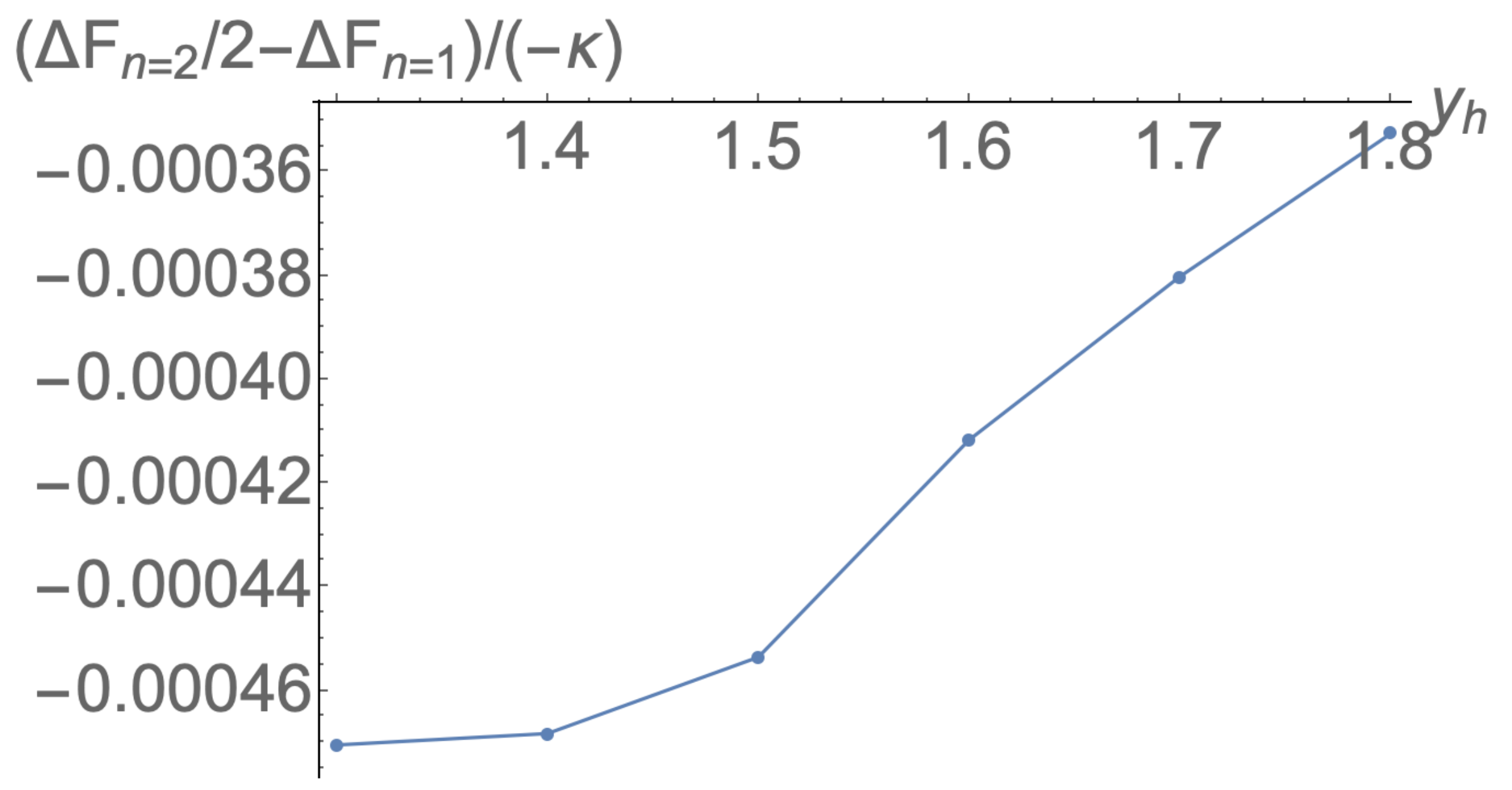}
\caption{$\chi=0$}
\end{subfigure}
\begin{subfigure}{.5\textwidth}
\centering
\includegraphics[width=\linewidth]{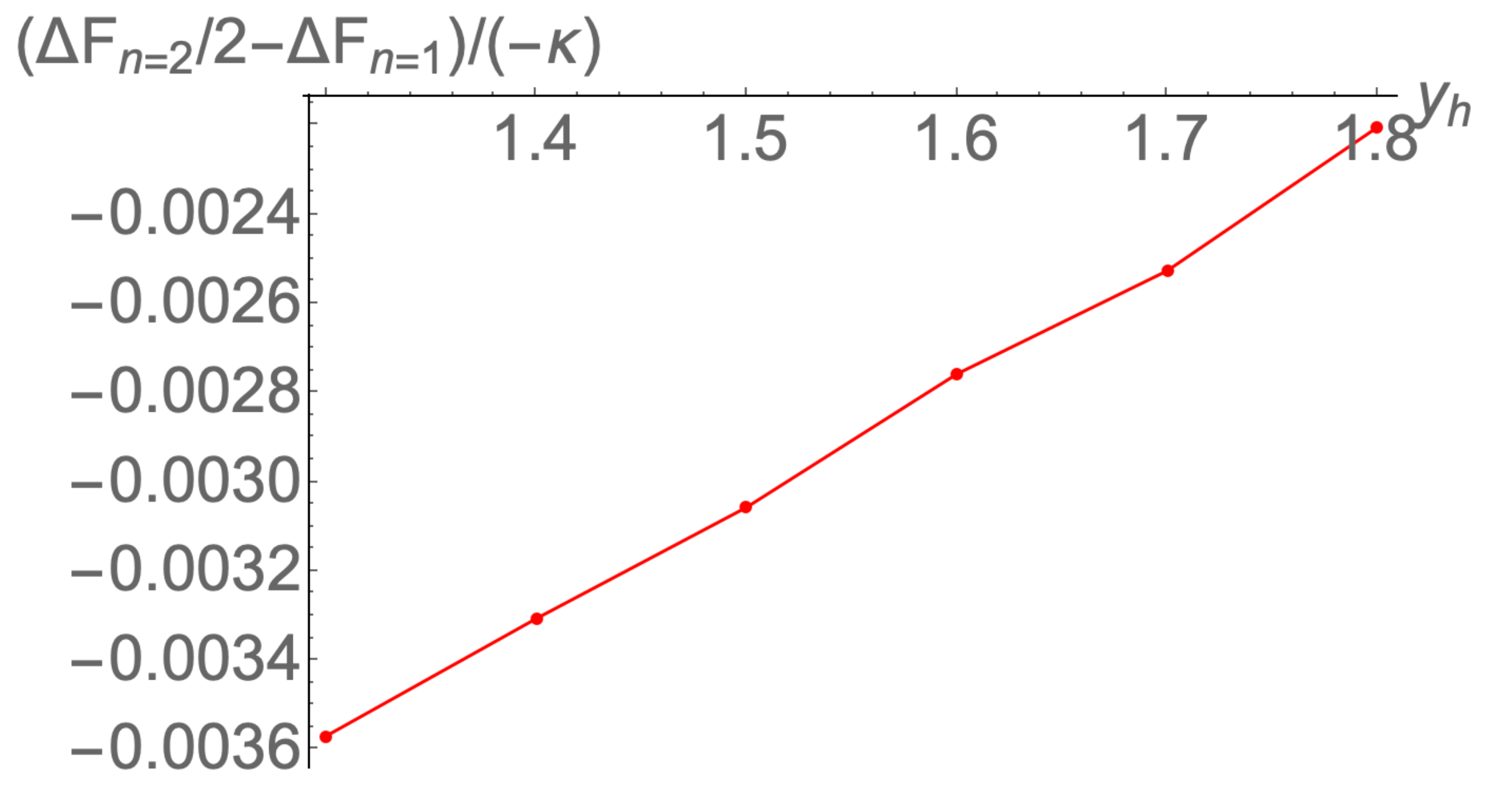}
\caption{$\chi\neq0$}
\end{subfigure}
\caption{Check of vortex type with same parameters as above, $qL =2$. For these parameters these solutions are of type I.}
\label{fig-free-energy-3}
\end{figure}

 For  all the temperatures and parameters  we scanned,
 we find (see figure  \ref{fig-free-energy}) that the phase with the $\chi$ condensate is preferred. 
This is analogous to what happens in flat space, where a similar model indicates 
non-abelian vortices to be energetically preferred over Abrikosov ones \cite{Shifman:2014oqa}. 
For all these solutions, we checked the first law $\Delta F = T \Delta S$ to be verified to order 0.1 $\%$ accuracy. 
This check would have failed if one did not use the conditions eq.(\ref{alfetta},\ref{rest}), 
due to the extra $\log \, y$ terms in the expansion.

Figure \ref{fig-free-energy-2} shows a comparison between the free energy 
between two winding $1$ vortices at very large distance and a
winding $2$ vortex, which corresponds to two coincident winding $1$ 
vortices. We find that the state with two far away vortices has 
a lower free energy (this is true both for the abelian vortex state and for the non-abelian
one, which has a $\chi$ condensate inside).
So we conclude that for $qL=1$ the non-abelian vortices are of type II. Similarly, we find that these non-abelian vortices are of type I for $qL = 2$, as shown in figure \ref{fig-free-energy-3}. \newline

A complete analysis of the phase diagram of these kind of vortices in flat space was performed in \cite{Tallarita:2017opp}, where explicit solutions for lattice configurations of non-abelian vortices were presented. A similar analysis would be interesting here and we leave this to future work.


\section{Vortex orientational zero modes}
\label{sect5}

The $SU(2)$ gauge symmetry is unbroken far away from 
the core of the non-abelian vortex discussed in section \ref{sect3},
but it is spontaneously broken nearby the center of the vortex at $x \rightarrow 0$.
In the boundary theory, this gives rise to classical Goldstone 
modes localized on the vortex world line.
Holographic Goldstone modes have been studied by
several authors, e.g. \cite{Amado:2009ts,Iqbal:2010eh,Anninos:2010sq,Amado:2013xya,Argurio:2015via}.
The situation that we study in this section  is different from these cases, 
because our symmetry breaking modes give rise to
a degree of freedom which is localized on a topological defect.

In principle, for a one-dimensional system, these modes should be
gapped from the Coleman-Mermin-Wagner theorem 
\cite{Coleman:1973ci,Mermin:1966fe}.
As described in \cite{Anninos:2010sq}, this kind of quantum effects in AdS
are subleading in the large $N_c$ expansion. The classical physics 
in the bulk gives a classical massless Goldstone in the boundary;
the description of the dynamics of the quantum system requires a 
one-loop calculation in the bulk. In this section we will study 
just the classical dynamics in the bulk, and we leave the 
quantum aspects for further work.

In order to study generic perturbations around the symmetry-breaking
non-abelian vortex solution, we introduce 
a unit 3-component vector $S^a$, which is an arbitrary function of the bulk coordinates, 
 which parameterize the $SU(2)$ orientation of the field $\chi^a$:
\be
\chi^a = S^a \chi_0 \, , \qquad S^a S^a = 1 \, ,
\ee
where we denote by $\chi_0$ the static non-abelian vortex profile,
determined in section \ref{sect3}. 
The profile function $\chi_0 $ is a function of  $(y,x)$ 
if we use the coordinated of the metric (\ref{metric}) or
 $(z,\tilde{x})$ if we use FG coordinates.
Replacing this ansatz in the action of the non-abelian sector eq. (\ref{esse-chi}),
we find the following effective lagrangian which describes
 perturbations from the non-abelian vortex solution:
\be
\label{eql1}
S_{\rm eff}=
\int d^4x \sqrt{-g}\left(\frac{\chi_0^2}{2} \hat{D}_\mu S^c\hat{D}^\mu S^c
-\frac{1}{4g_2^2} \tilde{F}_{\mu\nu}^a \tilde{F}^{a\mu\nu}\right).
\ee

If we consider small deviations from the non-abelian vortex solutions 
with $\chi^1=\chi^2=0$ discussed in section \ref{sect3},
it is useful to introduce the Goldstone fields $\pi_{1,2}$:
\beq
S^c=  \exp \le i \pi^a T^a \ri
\left( \begin{array}{c}
0 \\ 
0\\
1 \\
  \end{array}\right) \approx
  \left( \begin{array}{c}
\pi^1 \\ 
\pi^2  \\
1 \\
  \end{array}\right) \, , 
\eeq
where $T_a$ are the generators of the adjoint representation of $SU(2)$.
At quadratic order, the effective action is:
\bea
\label{eql1-linear}
S_{\rm eff}^q &=&
\int d^4x \sqrt{-g}  \left( -\frac{1}{4g_2^2} \tilde{F}_{\mu\nu}^a \tilde{F}^{a\mu\nu})  \right. \nl
&& \left. 
+\frac{\chi_0^2}{2} (\p_\mu \pi^1 -\tilde{A}^2_\mu) (\p^\mu \pi^1 -\tilde{A}^{2 \mu})
+\frac{\chi_0^2}{2} (\p_\mu \pi^2 -\tilde{A}^1_\mu) (\p^\mu \pi^2 -\tilde{A}^{1 \mu})
\right) \, .
\eea
At linear order, the equations of the scalar $\pi^1$ and gauge field $\tilde{A}_\nu^2$ then are
\be
\label{lingen1}
\partial_\mu\left(\chi_0^2 \sqrt{-g}g^{\mu\nu}(\partial_\nu\pi^1-\tilde{A}_\nu^2)\right)=0
\ee
\be\label{lingen2}
\frac{1}{4g_2^2}\frac{1}{\sqrt{-g}}\partial_\mu\left(\sqrt{-g}g^{\mu\alpha}g^{\nu\beta} 
\tilde{F}_{\alpha\beta}^2\right)+\frac{\chi_0^2}{2} g^{\mu\nu}(\partial_\mu \pi^1 -\tilde{A}_\mu^2)=0.
\ee
Analog equations can be written for $(\pi^2,\tilde{A}_\nu^1)$.
In the following we will solve the linear system in eqs. (\ref{lingen1},\ref{lingen2});
for simplicity we will use the following variables:
\beq
\pi=\pi^1 \, , \qquad H_\nu= \tilde{A}_\nu^2 \, , \qquad H_{\mu \nu}= \tilde{F}^2_{\mu \nu} \, .
\eeq
Note that at  linear order the non-abelian nature of the fields
is not important, because at this order $\pi$  and $H_\mu$ behave as abelian degrees of freedom.
In particular, in the asymptotic region where $\chi_0 \rightarrow 0$, 
the gauge field $H_\nu$ behaves in the linear approximation as an abelian
gauge field in asymptotically AdS spacetime, without symmetry breaking.

Solving the scalar  equation  (\ref{lingen1}) in the AdS background
(\ref{empty-ads}) and using the asymptotic behavior of $\chi$
in eq. (\ref{chi-asint}), which gives $\chi_0 \approx \a_3 z$ nearby the boundary
in the non-abelian vortex static solution, we find that 
for $z \rightarrow 0$, the Goldstone field satisfies $\partial^2_z \pi=0$.
This gives the asymptotic behavior of the field $\pi$ as:
\beq
\pi = B + A z + \dots \, .
\label{pipio}
\eeq
The asymptotic behavior of $\pi$  is then characterized by 
the two allowed values of $\Delta_\pi=0,1$; 
the condition for them to be normalizable is:
\beq
\Delta_\pi +\Delta_\chi \geq \frac{d-2}{2} \, , \qquad d=3 \, , \qquad \Delta_\chi=1 \, ,
 \eeq
 where the equality corresponds to the unitarity bound.
Both the solutions $\Delta_\pi=0,1$ are normalizable;
however, in order to enforce the Robin condition for $\chi$
(which in the boundary theory is dual to the double trace deformation
for the corresponding operator), we are forced to take $A$ as source
and $B$ as a VEV.

The asymptotics of the gauge field  $H_\nu$ is:
\beq
H_\mu = H_\mu^0 + H_\mu^1 z + \dots \, ,
\eeq
where in the following we will set the 
source $H_\mu^0=0$, and $H_\mu^1$ instead is proportional
to the expectation value of the current $J_\mu$.

\subsection{Consistency of equations}

Let us consider  eqs. (\ref{lingen1},\ref{lingen2}) more explicitly.
Before considering the Goldstone mode equations in the fully back-reacted
non-abelian vortex background, it is useful to consider 
eqs. (\ref{lingen1},\ref{lingen2}) in a probe limit, with the following 
diagonal metric:
\beq
ds^2=g_{tt} dt^2 +g_{zz} dz^2  + g_{\tx \tx} ( d \tx^2+ \tx^2 d \theta^2) \, ,
\eeq
where the diagonal metric coefficients are functions just of $z, \tx$.
A black brane solution can be recovered as a particular case, with the metric coefficients
depending only on $z$.
By gauge choice, we can set $H_z=0$; moreover, $H_\theta=0$ due to 
cylindrical symmetry.   We  consider a perturbation of the following form
for the remaining fields:
\beq
\pi = e^{-i \omega t} \hpi(z, \tx) \, ,   \qquad 
 H_t=e^{-i \omega t} \hH_t(z, \tx) \, , \qquad 
  H_{\tx}=e^{-i \omega t} \hH_x(z,\tx) \, .
  \label{omega-ansatz1}
\eeq
Eq.  (\ref{lingen1}) reads:
\beq
\chi_0^2  \sqrt{-g} g^{tt}  (-\omega^2 \hpi + i \omega \hH_t )  +
\p_z \le \chi_0^2  \sqrt{-g} g^{zz}  \p_z \hpi  \ri +
\p_\tx \le \chi_0^2  \sqrt{-g} g^{\tx \tx}  (\p_\tx \hpi - \hH_\tx )  \ri 
= 0 \, 
\label{probe0}
\eeq
The $\nu=z,t,\tx$ components of eq. (\ref{lingen2}) give respectively the following equations:
\beq
  i \omega g^{tt} g^{zz}   \p_z \hH_t   
  -  \frac{ \p_\tx ( \sqrt{-g} g^{\tx \tx} g^{zz}  \p_z \hH_\tx )  }{  \sqrt{-g}} + 
4 g^2_2 \frac{\chi_0^2}{2} g^{zz}  \p_z \hpi    = 0 \, , 
\label{probe1}
\eeq
\beq
\frac{ \p_z (  \sqrt{-g} g^{zz} g^{tt} \p_z \hH_t ) + 
 \p_\tx ( \sqrt{-g} g^{\tx \tx} g^{tt} ( \p_\tx \hH_t + i \omega \hH_\tx  ) )  }{  \sqrt{-g}}+ 
4 g^2_2  \frac{\chi_0^2}{2} g^{tt} (-i \omega \hpi - \hH_t )  = 0 \, , 
\label{probe2}
\eeq
\beq
\frac{ \p_z (  \sqrt{-g} g^{zz} g^{\tx \tx} \p_z \hH_\tx )  }{  \sqrt{-g}}
- i \omega g^{tt} g^{\tx \tx}  ( - \p_\tx \hH_t -i \omega \hH_\tx )
 + 4 g^2_2  \frac{\chi_0^2}{2} g^{\tx \tx} (\p_\tx \hpi - \hH_\tx )  = 0 \, , 
 \label{probe3}
\eeq
The $\nu=\theta$ component of (\ref{lingen2}) gives instead a trivial equation.
If one solves eqs. (\ref{probe1},\ref{probe3}) for $(\p_z \hpi, \p_\tx \hpi)$
and replaces it in eq. (\ref{probe0}), one finds eq. (\ref{probe2}).
So eq. (\ref{probe2}) is redundant, and indeed in this way we have
three independent differential equations for the three unknown functions
$(\hpi, \hH_\tx, \hH_t)$.

These considerations extend in a direct way  to the case with back-reaction.
In this case, using the coordinates $(t,y, x,\theta)$ one can consider the
general metric with cylindrical symmetry
\beq
g_{\mu \nu}=\left(
\begin{array}{cccc}
 g_{tt}&   &  &  \\
 &  g_{yy} &   g_{yx} &  \\
  &  g_{yx} & g_{xx}   &  \\
    &  &   & g_{\theta \theta}  \\
  \end{array}\right) \, ,
  \label{mmetr}
  \eeq
 where each of the metric coefficient is a function of $y,x$.
 The metric (\ref{metric}) used in numerical calculations is just a different parameterization
 of (\ref{mmetr}).
   By gauge choice, we can set $H_y=0$; moreover, $H_\theta=0$ due to 
cylindrical symmetry.   We consider the following perturbation
for the remaining fields:
\beq
\pi = e^{-i \omega t} \hpi(y, x) \, ,   \qquad 
 H_t=e^{-i \omega t} \hH_t(y, x) \, , \qquad 
  H_{x}=e^{-i \omega t} \hH_x(y,x) \, .
  \label{ansazzalo}
\eeq
The $\theta$ component of eq. (\ref{lingen2}) gives again a trivial equation.
One can solve for $(\p_y \hpi, \p_x \hpi)$
using the $(z,x)$ components of eq. (\ref{lingen2}).
If we replace these expression in eq. (\ref{lingen1}),
one finds (after cumbersome calculations) an equation which is equivalent 
to $t$ component  of eq. (\ref{lingen2}).
So again we have  three independent equations
for the three unknown functions $(\hpi, \hH_x, \hH_t)$,
which is again a  well-posed problem.

\subsection{Two-point functions}

We will study the two-point functions for the Goldstone
field $\pi$ and we will ignore the boundary terms coming from the gauge fields. These Goldstone fields correspond to the gapless orientational moduli of the dual vortex solution.
After integration by parts, and using the equations of motion, 
the action (\ref{eql1-linear}) has a contribution coming from a boundary term:
\be
S^q_b=-\int d^3 w \, \sqrt{-h} \chi_0^2 \pi n^\mu \hat{D}_\mu\pi = 
-\int d^3w \, \sqrt{-h} \chi_0^2 \pi n^\mu \partial_\mu \pi,
\ee
where $w^A=(t, \tx, \theta)$ are the boundary FG coordinates,
$n^\mu=(z/L,0,0,0)$ is the unit normal to the boundary 
 and we have fixed the gauge so that $H_z =0$. 

Nearby the boundary, the profile function $\chi_0$ has the following form:
\be
\chi_0 = \a_0 z + \b_0 z^2 + \dots \, , 
\quad 
\chi_0^2 = \a^2_0 \, z^2  + 2 \a_0 \b_0 \, z^3 + \dots \, ,
\ee
where $\a_0$ and $\b_0$ are functions of $\tx$,
with $\b_0= \eta \a_0$, due to the Robin condition.
The bulk Goldstone field $\pi$ instead, near the boundary has the form given
in eq. (\ref{pipio}), where $B$ and $A$ are functions of $(\tx, t)$.

Expanding the action around the boundary we get 
\be
S^q_b= -\int d^3w \,  \frac{L^2}{z^2} \chi_0^2 \pi \partial_z \pi
 =- \int d^3w \,  L^2\left(  \a_0^2 A B+z \left( \a_0^2 A^2 +2 \a_0 \b_0 AB\right)+O(z^2)\right) \, .
\ee
In order to enforce the Robin condition for the field $\chi$, 
in the boundary we are forced to  treat $A$
as a source and $B$ has a VEV.

We expand the time dependence in Fourier series as in eq. (\ref{omega-ansatz1})
and we consider the profiles near the boundary:
\beq
\hpi = B_\omega(\tx) + A_\omega(\tx) z + \dots \, .
\eeq
We will consider situations where the source $A_\omega$ 
is independent of the radial boundary coordinate $\tx$, this can always be enforced by a suitable boundary condition.

Solving for the equations of motion (\ref{lingen1},\ref{lingen2}), 
and differentiating $S^q_b$ with respect to $A_\omega$,
we obtain a VEV which is a function of $\tx$;
normalizing with respect to the source, we obtain the two point function of the operator:
\be
\big<\mathcal{O}_B (\tx,\omega)\mathcal{O}_B (0,\omega)\big> 
=-\frac{ B_\o(\tx)(\a_0 (\tx))^2}{A_\omega} = \xi_B(\tx, \omega) \,.
\ee
Integrating this in the spatial coordinate $\tx$, we obtain the averaged two-point function $ \hat{\xi}_B$,
defined on the vortex world line:
\be
\hat{\xi}_B(\omega) = \int_0^1 \frac{\tx}{(1-\tx)^3} \xi_B(\tx, \omega) d \tx \, .
\ee
The energy dissipated by the source per unit of time is proportional to:
\be
\frac{dE}{dt} \propto \o  A_\o^2 \, \im \,  \hat{\xi}_B(\omega) \, .
\ee

\subsection{Small $\o$ limit}  
\label{subsect:small-omega}
  
For $\omega=0$  eqs. (\ref{lingen1},\ref{lingen2})  are all solved by
 a constant $\hpi(y,x)=\hpiz$ and a zero $\hH_\mu$. 
By gauge invariance, one can perform a gauge transformation
$\pi \rightarrow e^{-i \omega t} \pi$, but this condition
induces a non zero value for the source of the 
symmetry current $H_\mu^0$, and so it is dual to a non zero
chemical potential on the boundary theory.
In order to study non-zero $\omega$, we need then to solve the coupled 
system of differential equations (\ref{lingen1},\ref{lingen2}).   
Let us consider the small $\omega$ limit
of eqs. (\ref{probe1}-\ref{probe3}) in the probe limit 
(the back-reacted case is qualitatively similar).
It turns out that the following expansion in $\omega$ is consistent with the
equations of motion:
\beq
\hpi=\hpiz+\omega^2 \hpiq \, , \qquad \hH_t=\omega {\hHl_t} \, ,
\qquad \hH_x=\omega^2  {\hHq_x} \, .
\eeq
The function $\hHl_t$ can be found solving eq. (\ref{probe2}), which in the small $\o$ limit reads:
\beq
\frac{ \p_z (  \sqrt{-g} g^{zz} g^{tt} \p_z \hHl_t ) + 
 \p_\tx ( \sqrt{-g} g^{\tx \tx} g^{tt} ( \p_\tx \hHl_t  ) )  }{  \sqrt{-g}}+ 
2 g^2_2  \chi_0^2 g^{tt} (-i  \hpiz - \hHl_t )  = 0 \, . 
\eeq
Taking $\hpiz$ real, this gives an imaginary $\hHl_t$.
The functions  $\hpiq,\hHq_t$ can be found from eqs. (\ref{probe1},\ref{probe3}), which in 
the small $\o$ limit give:
\beq
  i  g^{tt} g^{zz}   \p_z \hHl_t   
  -  \frac{ \p_\tx ( \sqrt{-g} g^{\tx \tx} g^{zz}  \p_z \hHq_\tx )  }{  \sqrt{-g}} + 
2 g^2_2 \chi_0^2 g^{zz}  \p_z \hpiq    = 0 \, , 
\eeq
\beq
\frac{ \p_z (  \sqrt{-g} g^{zz} g^{\tx \tx} \p_z \hHq_\tx )  }{  \sqrt{-g}}
+ i  g^{tt} g^{\tx \tx}   \p_\tx \hHl_t 
 + 2 g^2_2  \chi_0^2 g^{\tx \tx} (\p_\tx \hpiq - \hHq_\tx )  = 0 \, , 
\eeq
This equations give a real $\hpiq$ and $\hHq_x$, while
 the imaginary part of $\hpi$ starts at order $\o^3$.

From these small $\o$ solutions, we find:
\beq
 B_\o \propto \o^0 \, , \qquad A_\o \propto \o^2 + i \Upsilon \,  \o^3  \, ,
\eeq
where the leading parts in $\o$ are real and the imaginary part, 
whose coefficient is denoted by $\Upsilon$, is subleading.
These considerations give that at small $\omega$:
\beq
\re \, \hat{\xi}_B \propto 1/ \omega^2 \, , \qquad
\im \, \hat{\xi}_B \propto 1/ \omega \, ,
\eeq
which gives the pole for the Goldstone mode in the 
dual boundary theory. 
We will now solve these equations numerically to find these modes explicitly and verify these behaviours.

\subsection{Numerical calculations}
  
 In order to solve eqs. (\ref{lingen1}, \ref{lingen2}) numerically, 
we use Eddington-Finkelstein (EF) coordinates, 
which are the natural coordinates to describe infalling boundary conditions on the black hole horizon. 
In order to pass the metric (\ref{metric}) to EF coordinates, we define a shifted time $v$ such that
\be
dt = dv - \frac{dy}{1-y^3}\sqrt{\frac{Q_2}{Q_1}}\frac{1}{y_+}
\ee
which can be obtained form radial null geodesics. 
In these coordinates then the metric is:
\bea
\label{metricEF}
ds^2_{EF} &=& \frac{L^2}{y^2}\left\{ -Q_1 y_+^2(1-y^3)dv^2
+2 dv dy \left(y_+\sqrt{Q_1Q_2}\right) \right. \nn\\
&& \left.  +\frac{y_+^2Q_4}{(1-x)^4}(dx+xy^2(1-x)^3Q_3 dy)^2 + \frac{y_+^2 Q_5 x^2}{(1-x)^2}d\theta^2 \right\} \, ,
\eea
which is smooth at $y=1$.

The boundary condition for the field $\hH_x, \hH_t$ are set us follows:
\begin{itemize}
\item The sources for the $SU(2)$ symmetry are set to zero, i.e.
at $y \rightarrow 0$, we set $\hH_x=\hH_t=0$. 
\item Ar $x \rightarrow 1$, we should impose that 
 the gauge field $H_\mu$ is zero far away the vortex
core (where $\chi_0=0$).
This can be imposed setting
\beq
\hH_x=\hH_t=0 \, , \qquad \p_x \hH_x=\p_x \hH_t=0 \, ,
\eeq
for every $y$ at $x \rightarrow 1$.
\item At $y \rightarrow 1$ we set that the gauge field is smooth in EF coordinates
\item At $x \rightarrow 0$, due to cylindrical symmetry,
we should have: $\p_x \hH_t=\p_x \hH_x =0$.
\end{itemize}

Then we can solve eqs. (\ref{lingen1}, \ref{lingen2}) in the background of the back-reacted
non-abelian vortex solution that we have found in sect \ref{sect3}. 
Therefore we input the $Q_i$ and the $\chi_0$ solutions we have obtained before 
and solve for $\pi$ and $H_\mu$ to extract the two-point function of $\pi$. 
 The ansatz for the fields $\pi, H_\mu$ is taken as in eq. (\ref{ansazzalo}).   
We fix a constant value of the source $A_\o=1$ at the boundary, 
which correspond to a value of $\partial_y \pi$ independent from $x$.

\begin{figure}[ptb]
\begin{subfigure}{.5\textwidth}
\centering
\includegraphics[width=0.9\linewidth]{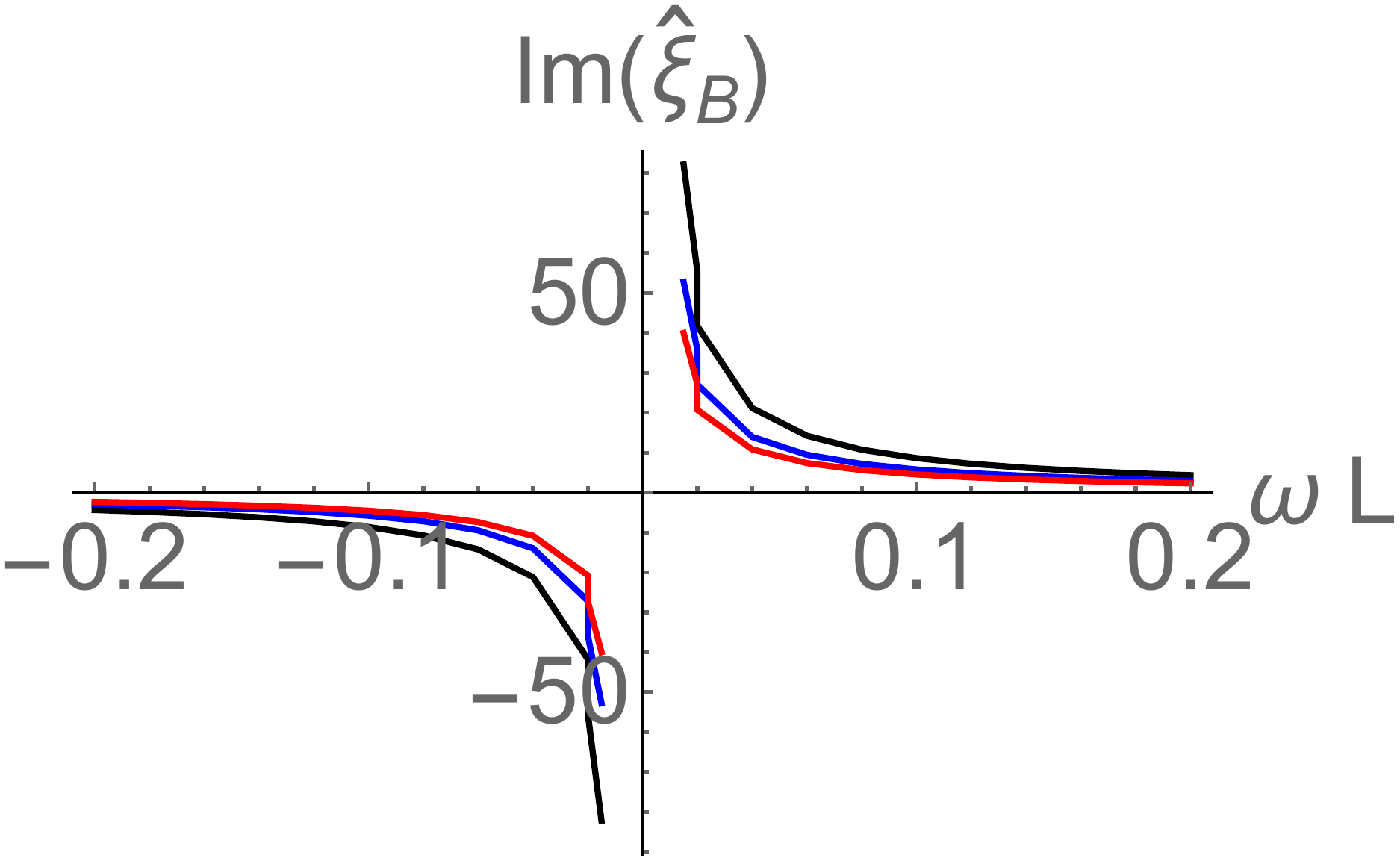}
\end{subfigure}
\begin{subfigure}{.5\textwidth}
\centering
\includegraphics[width=0.9\linewidth]{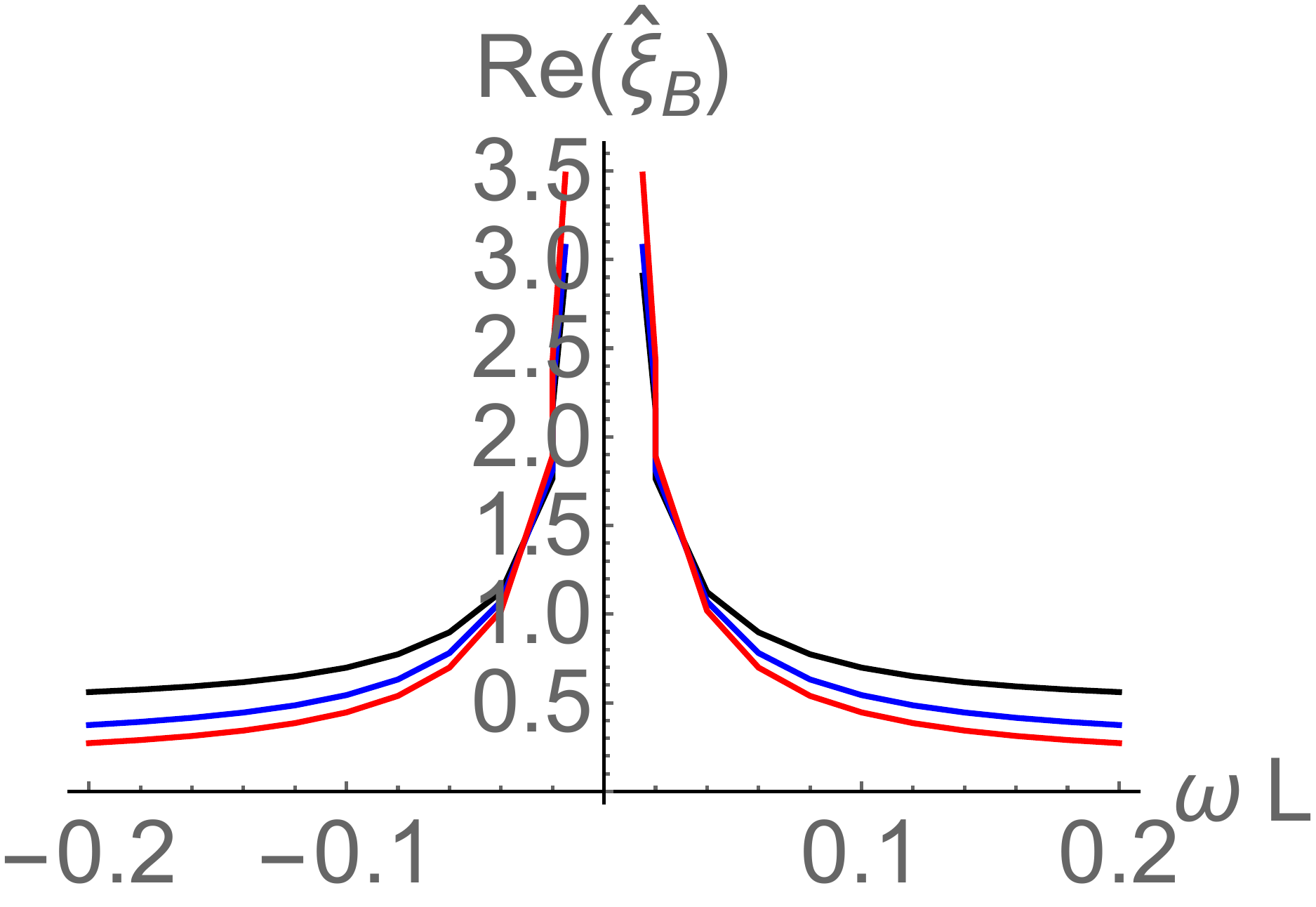}
\end{subfigure}
\caption{Solutions of the integrated two point function $\hat{\xi}_B(\o)$ 
of the orientational perturbations at varying temperatures. 
These are $y_+ =2 \text{(red line)}, 3/2 \text{(solid blue line)}, 1 \text{(dashed line)}$.}
\label{fig6}
\end{figure}

We compute the function $\xi_B(x)$ solving the equations of motion;
the dependence of the solution from  $x$ is rather weak.
Solutions for the averaged $\hat{\xi}_B$ are shown  in figure \ref{fig6}
 for varying temperatures (the plot for $\xi_B(x=0)$ is almost identical).
  The real part  $\re \, \hat{\xi}_B$
 scales as $1/\o^2$ at small $\o$ and  the imaginary part $\im \, \hat{\xi}_B$ scales as $1/\o$:
  this is consistent  with the discussion in section \ref{subsect:small-omega}.
 Therefore, we have determined the presence of classically gapless degrees of freedom on the dual vortex, 
 corresponding to its orientational moduli.

\section{Conclusions}
\label{sect-conclu}

In this paper we studied a model of non-abelian vortex in 
an asymptotically AdS$_4$ spacetime. This system is dual to 
a deformed CFT in $2+1$ dimensions
 at finite temperature in the presence of a vortex point-like defect.
 We computed the back-reaction of the vortex solution on 
the geometry, and we found a regime of parameters in
which the free energy of the non-abelian vortex is lower than
the free energy of the $U(1)$ vortex. This 
provides the first explicit realisation of
a non-abelian vortex in a gravity dual which includes
the vortex gravity back-reaction. 
We found that, depending on charge of the abelian field $q$,
these vortices can be type I or type II.

In the example that we studied, the non-abelian vortex has a $\mathbb{CP}^1$ zero mode
which, from the bulk point of view, is localized on a string worldsheet;
in the boundary dual field theory it corresponds to a $\mathbb{CP}^1$
quantum rotor interacting with a strongly coupled CFT at finite temperature.
At the classical level in the bulk,
which corresponds to leading order in the number of CFT degrees of freedom (e.g. number of colors $N_c$), 
the zero modes give Goldstone bosons localized on the  soliton
world line.  We have computed the two-point functions of these
zero modes and we found a pole corresponding to a classically
massless degree of freedom.  

 The spectrum of excitations localized on the vortex should be gapped 
due to the Coleman-Mermin-Wagner theorem 
\cite{Coleman:1973ci,Mermin:1966fe}.
This is a subleading effect in the number of colours $N_c$
and should be described by bulk quantum corrections,
in analogy to the case  of Goldstone zero-modes localized 
in the full boundary theory \cite{Anninos:2010sq}.
Note that the large $N$ is not from the point of view of the vortex
worldsheet, which is always described by a  $\mathbb{CP}^1$ sigma model,
but rather the large $N_c$ limit is in the CFT to which the defect is coupled.
There is no decoupling between the infrared of the worldsheet vortex
theory and the CFT degrees of freedom, because there is no mass gap
in the field theory. This feature should give a suppression of the mass
gap on the vortex world volume, which somehow disappears 
as a leading order effect in the gravity dual.
It would be interesting to generalise the analysis of \cite{Anninos:2010sq}
to the case of Goldstone bosons localized on a topological defect.
We leave this problem for future investigation.

Non-abelian vortices can be realised as a probe D-brane in 
the Polchinski-Strassler \cite{Polchinski:2000uf}
 gravity dual of $\mathcal{N}=1^*$ theory \cite{Auzzi:2008ep}
and of mass-deformed ABJM theory \cite{Auzzi:2009es}.
In these cases, as in the present one, the mass gap on top of the non-abelian vortex
world volume theory (which is a $\mathbb{CP}^1$ sigma model for every number
of colors) does not come directly from the gravity dual, but requires
quantization of the worldsheet theory (which comes from the D-brane
Dirac-Born-Infeld action).  In this paper we focused on a simpler
bottom-up realisation of a non-abelian $\mathbb{CP}^1$ 
vortex in holographic dual; the outcome is still that the mass gap
on the vortex worldsheet  is suppressed in the $1/N_c$
expansion of the boundary CFT, and so it corresponds to quantum
effects in the bulk.

 \section*{Acknowledgments}

We are grateful to Oscar Dias and Jorge Santos for useful discussion
of  technical details in ref. \cite{Dias:2013bwa}.
R. A. and A. P. thank the  Departamento de Ciencias of
 Universidad Adolfo Ib\'a\~nez for hospitality when this work was in progress.  This work was funded by the NSERC Discovery Grant (A.P.). G.T. is funded by a Fondecyt grant number 11160010.

\appendix

\section*{Appendix A}
\label{AppeA}

The followings series expansions of $Q_k$ are fixed by the equation of motion:
\bea
Q_1^{(1)}&=& - \, Q_2^{(1)}=Q_4^{(1)}=Q_5^{(1)}=\frac{5 \ta }{8} \, ,
\nl
Q_3^{(1)} &=&
-\frac{(x-1) \left(2 n x^{2 n} Q_6^{(0)}(x){}^2+2 x^{2 n+1} Q_6^{(0)}(x)
   {Q_6^{(0)}}'(x)+x Q_8^{(0)}(x) {Q_8^{(0)}}'(x)\right)}{4 x^2 y_+^2} \, ,
\nl
Q_1^{(2)}&=&Q_4^{(2)}=Q_5^{(2)}=
\frac{1}{512} \left(215 \ta ^2-256 \ta -256 x^{2 n} {Q_6(x)^{(0)}}^2-128
    {Q_8(x)^{(0)}}^2\right) \, ,
\nl
Q_2^{(2)}&=&\frac{1}{128} \left(128 \ta -45 \ta ^2\right) \, ,
\eea

\beq
Q_2^{(3)}= \frac{
26880 y_+ \ta^2 -7725 y_+ \ta^3 + x^{2 n} {Q_6^{(0)}}^2 (7936 y_+ \ta +32768 \kappa_1)
+ {Q_8^{(0)}}^2 (3968 y_+ \ta + 16384 \kappa_2)}{24576 y_+} \, ,
\eeq
\beq
Q_1^{(3)}+Q_4^{(3)}+Q_5^{(3)}=\frac{75 \ta ^2 (109 \ta -256)}{8192}
-\frac{2 x^{2n} (37 y_+ \ta +128 \kappa_1) {Q_6^{(0)}}^2+
 (37 y_+ \ta +128 \kappa_2) {Q_8^{(0)}}^2}{64 y_+} \, ,
\eeq
\beq
Q_5^{(3)}-Q_4^{(3)}=
\frac{(1-x)x}{96 y_+}
\le  (160 \kappa _1+17 \ta  y_+)   [(x^2 Q_6^{(0)})^2]' 
+(160 \kappa _2+17 \ta  y_+)  \frac{[( Q_8^{(0)})^2]'}{2}  
+96 y_+ (Q_4^{(3)})'
\ri
\eeq
These coefficients are symmetric
under exchange of $x^n Q_6^{(0)} $ and $ \frac{Q_8^{(0)}}{\sqrt{2}} $, exchanging
also $\kappa_1$ and $\kappa_2$.

The coefficients of the expansion for the transformation of the metric
to FG coordinates are:
\bea
a_2&=&\frac{5 \ta  y_+^2}{16} \, , \qquad 
a_3= \frac{\ta  (265 \ta -256) y_+^3}{1024} \, , \nl
a_4&=&\frac{y_+^4 \left(6025 \ta ^3-8960 \ta ^2-4096 Q_2^{(3)}-4096\right)}{24576} \, , 
\nl
b_1&=&b_2=b_3=0 \, , \qquad b_4=-\frac{1}{4}  (1-x)^3 x y_+^4  Q_3^{(1)} \, ,
\eea

\end{document}